\documentclass[journal]{IEEEtran}
\usepackage{xcolor,soul,framed} 
\usepackage{cite}
\usepackage{dsfont}
\usepackage{amssymb}
\usepackage{upgreek}
\usepackage{caption}
\usepackage{subcaption}
\colorlet{shadecolor}{yellow}
\usepackage[pdftex]{graphicx}
\graphicspath{{../pdf/}{../jpeg/}}
\DeclareGraphicsExtensions{.pdf,.jpeg,.png,.bmp}
\usepackage{bm}
\usepackage{lipsum, color}
\usepackage[cmex10]{amsmath}
\usepackage{color, colortbl}
\definecolor{Gray}{gray}{0.9}
\definecolor{LightCyan}{rgb}{0.88,1,1}
\usepackage{nccmath}
\usepackage{eqparbox}
\usepackage{url}
\usepackage{amsfonts,amsthm}
\usepackage[utf8]{inputenc}
\usepackage[english]{babel}
\usepackage{multirow}

\DeclareMathOperator{\E}{\mathbb{E}}                     
\DeclareMathOperator{\Var}{\mathbb{V}ar}
\DeclareMathOperator{\erf}{\text{erf}}
\DeclareMathOperator{\SNR}{\mathtt{SNR}}

\makeatletter
\let\saveqed\qed
\renewcommand\qed{%
   \ifmmode\displaymath@qed
   \else\saveqed
   \fi}

\makeatletter
\newtheoremstyle{myStyle}
  {1ex plus 0.2ex minus 0.1ex} 
  {1ex plus 0.2ex minus 0.1ex} 
  {\itshape}    
  {}            
  {\bfseries}   
  {.}           
  {0.5em}       
  {\thmname{#1}~\thmnumber{#2}~\thmnote{(#3)}}
\makeatother
\theoremstyle{myStyle}

\newtheorem{lemma}{Lemma}
\newtheorem{proposition}{Proposition}

\usepackage{mathtools, cuted}
\usepackage{lipsum, color}
\usepackage{lipsum}
\usepackage{mathtools}
\usepackage{cuted}
\usepackage{flushend}
\usepackage{silence}
\usepackage{etoolbox}
\tracingpatches
\makeatletter
\usepackage{amsthm}
\usepackage{tabularx}
\newcolumntype{P}[1]{>{\centering\arraybackslash}p{#1}}
\allowdisplaybreaks
\newcommand{\tcb}[1]{\textcolor{black}{#1}}

\begin{document}
\bstctlcite{IEEEexample:BSTcontrol}
    \title{STBC-Aided Cooperative \textcolor{black}{NOMA} 
    with Timing Offsets, Imperfect \textcolor{black}{Successive Interference Cancellation, and Imperfect Channel State Information}}
  \author{Muhammad Waseem~Akhtar,~\IEEEmembership{Student Member,~IEEE,}  
  Syed Ali~Hassan,~\IEEEmembership{Senior Member,~IEEE,} Sajid~Saleem,~\IEEEmembership{Member,~IEEE, }Haejoon~Jung,~\IEEEmembership{Member,~IEEE}
  \thanks{M. W. Akhtar, S. A. Hassan, and S. Saleem are with the  School of Electrical Engineering and Computer  Science~(SEECS), National University of Sciences and Technology~(NUST), Islamabad, Pakistan. (e-mail: engr.waseemakhtar@seecs.edu.pk,   ali.hassan@seecs.edu.pk, sajid.saleem@seecs.edu.pk.)

  H. Jung is with the Department of Information and Telecommunication Engineering, Incheon National University, Incheon  22012, Korea (e-mail: haejoonjung@inu.ac.kr)}
   }
 \maketitle
 
\begin{abstract}
 \textcolor{black}{
The combination of non-orthogonal multiple access (NOMA) and cooperative communications can be a suitable solution for fifth generation~(5G) and beyond 5G (B5G) wireless systems with massive connectivity, because it can provide higher spectral efficiency, lower energy consumption, and improved fairness compared to the non-cooperative NOMA.} However, the receiver complexity in the conventional cooperative NOMA increases with increasing number of users owing to \textcolor{black}{successive interference cancellation}~(SIC) at each user. Space time block code-aided cooperative NOMA (STBC-CNOMA) offers less numbers of SIC as compared to that of conventional cooperative NOMA. 
\textcolor{black}{
In this paper, we evaluate the performance of STBC-CNOMA under practical challenges such as imperfect SIC, imperfect timing synchronization between distributed cooperating users, and imperfect channel state information~(CSI). We derive closed-form expressions of the received signals in the presence of such realistic impairments and then use them to evaluate outage probability. Further, we provide intuitive insights into the impact of each impairment on the outage performance through asymptotic analysis at high transmit signal-to-noise ratio.} We also compare the complexity of STBC-CNOMA with existing cooperative NOMA protocols for a given number of users. \textcolor{black}{In addition, through analysis and simulation, we observe that the impact of the imperfect SIC on the outage performance of STBC-CNOMA is more significant compared to the other two imperfections. Therefore, considering the smaller number of SIC in STBC-CNOMA compared to the other cooperative NOMA protocols, STBC-CNOMA is an effective solution to achieve high reliability for the same SIC imperfection condition.}

\end{abstract}
\IEEEpeerreviewmaketitle

\section{Introduction}
\IEEEPARstart{N}{on}-orthogonal multiple access (NOMA) is considered to be one of the most promising techniques for fifth-generation (5G) and beyond 5G (B5G) wireless systems to meet the heterogeneous demands on low latency, high reliability, massive connectivity, improved fairness, and high throughput~\cite{sur3}. The key principle behind NOMA is to exploit non-orthogonal resource allocation among multiple users at the cost of increased receiver complexity, which is required for separating the non-orthogonal signals~\cite{Survey2}. In contrast to orthogonal multiple access~(OMA), multiple users in NOMA are assigned the same physical resource~(e.g., frequency and time) but with different power, which significantly enhances spectral efficiency.


 
Motivated by such advantage of NOMA, various aspects of NOMA have been actively investigated, engaging industry, standardization bodies, and academia. Further, as noted in~\cite{Survey3}, NOMA can be flexibly combined with various existing and emerging wireless technologies. 
In particular, the combination of NOMA and cooperative communications can be a suitable solution for the Internet-of-Things (IoT) networks with massive connectivity, because it can provide higher spectral efficiency, lower energy consumption, and improved fairness compared to the non-cooperative NOMA~\cite{Survey4}.

\tcb{In one of the} pioneering studies on the NOMA schemes that \tcb{incorporates the principles of cooperative communications}, the authors in~\cite{ZDCNOMA} propose \emph{cooperative NOMA}, which is \textcolor{black}{subsequently referred} to as conventional cooperative NOMA~(CCN).
In this scheme,  \emph{strong} users with better channel conditions support \emph{weak} users with worse channel conditions by serving as relays, which increases the reliability of the weak users through diversity gain.
In~\cite{complx1}, cooperative NOMA \tcb{is combined with} simultaneous wireless information and power transfer (SWIPT) to improve energy efficiency through energy harvesting. Further, in~\cite{complx3}, full duplex relaying-based NOMA schemes are introduced to reduce the \tcb{number} of time slots \tcb{required} to relay weak users' messages. Similarly, cooperation among users by means of full-duplex device-to-device~(D2D) communication is discussed  in~\cite{04}, where the outage performance of weak users is enhanced with the assistance of the full-duplex relaying by strong users. In addition, the authors in~\cite{01} propose a two-stage relay selection scheme for cooperative NOMA, which also provides lower outage rates. NOMA techniques adopting cooperative relaying systems are also extensively studied. For example, in~\cite{CRS-NOMA}, the authors propose an algorithm called cooperative relaying system  using NOMA (CRS-NOMA), in which a decode-and-forward (DF) relay is adopted. Also, assuming a single DF relay and two far users, the outage performances of different relaying schemes are investigated in~\cite{liu2018decode}. Also, NOMA using an amplify-and-forward (AF) protocol is investigated in~\cite{03}. 

Despite their effectiveness, the aforementioned NOMA schemes combined with cooperative communications incur an excessive number of successive interference cancellation (SIC) executed at user \tcb{terminals as} compared to the non-cooperative NOMA~\cite{jamal2018efficient}. When it comes to the IoT networks with limited capabilities (e.g., computational resources and power), users may suffer from prohibitively large energy consumption due to the excessive number of SIC. To overcome this issue, space time block code (STBC)-aided cooperative NOMA protocols are proposed, which benefit from diversity gain with reduced number of SIC. For instance, the conventional Alamouti (i.e., $2 \times 1$) STBC-based NOMA system is investigated in~\cite{NOMA_STBC}, which uses two antennas at the BS and a single antenna at each user. This scheme doubles the diversity order as compared to that of conventional NOMA. Furthermore, in~\cite{kader2016cooperative}, the authors propose an Alamouti STBC-based CRS-NOMA protocol for a network with source, relay, and destination, which are equipped with two transmit antennas, two transmit and one receive antennas, and one receive antenna, respectively. It shows higher sum capacity and lower outage probability compared to the conventional CRS-NOMA in~\cite{CRS-NOMA}. Instead of using the co-located (or real) antenna array, the authors in~\cite{jamal2017new, jamal2018efficient} propose a new cooperative NOMA with a \emph{distributed} STBC (i.e., STBC-CNOMA) for the virtual antenna array created by a group of single antenna users, which can be readily applied to the IoT networks. In their proposed scheme, STBC-CNOMA, they employ $2 \times 2$ distributed STBC on the NOMA system, in which two strong users act as DF relays and transmit the messages of the weak users by a $2 \times 2$ STBC. They show that STBC-CNOMA can achieve higher throughput with smaller number of SIC compared to the CCN in~\cite{ZDCNOMA}.

However, some challenges need to be addressed in the STBC-CNOMA systems in practical scenarios. For instance, distributed nature of terminals and their mobility cause the timing offsets, which is especially severe in virtual antenna array-based approaches including distributed STBC~\cite{23, 24}. In addition, reliability performance of NOMA can be significantly degraded by imperfect SIC, as reported in~\cite{Imperfect_SIC}, \textcolor{black}{and imperfect CSI~\cite{ipCSI1,ipCSI2,ipCSI3}}. However, the existing studies on STBC-CNOMA including~\cite{jamal2018efficient} and~\cite{jamal2017new} assume the ideal case without considering such realistic impairments. For this reason, in this paper, to better evaluate STBC-CNOMA, we investigate the impacts of the timing offsets, imperfect SIC, and \textcolor{black}{imperfect CSI} on its performance.
Our main contributions can be summarized as follows.

\begin{itemize}
	\item \textcolor{black}{To the best of our knowledge, it is the first comprehensive study on practical impairments in STBC-CNOMA. We present a theoretical framework including signal model of an arbitrary user under the timing mismatch, imperfect SIC, and channel estimation error.
	}
	
	\item \textcolor{black}{We derive the probability distributions of the signal-to-interference-plus-noise ratio (SINR) for different combinations of the three impairments, which can be used in the STBC-CNOMA system design and operation. Based on the derived distributions, we also provide the closed-form expressions of outage probabilities, which are not present in prior arts. 
	}
	
	\item \textcolor{black}{We also provide asymptotic rate (or capacity) outage probability in the high transmit signal-to-noise ratio (SNR) regime, which offers intuitive insights into how each of the three impairment hurts the performance of STBC-CNOMA.
	}
	
	\item \textcolor{black}{For the fair comparison with other cooperative NOMA protocols, we quantify the total number of SIC, the total number of required time slots, and the total number of transmissions of STBC-CNOMA with four existing schemes as functions of the number of user terminals.
	}
	
	\item \textcolor{black}{Numerical and simulation results are presented with different degrees of the three impairments. Through the comparison between analysis and simulation, we validate our analysis on the SINR distribution, exact and asymptotic capacity outage probabilities. In addition, we compare the outage performance of STBC-CNOMA with  CCN and non-cooperative NOMA. 
	}
\end{itemize}

The rest of the paper is organized as follows. In Section~\ref{sec:system_model}, we introduce the system model. \textcolor{black}{Three practical impairments~(i.e.,  timing error, imperfect SIC and imperfect CSI)} and corresponding signal models are presented in Section~\ref{sec:three_practical_impairments}. Section~\ref{sec:outage_probability_analysis} provides closed-form expressions of outage rates both in the absence and in the presence of the imperfections. Furthermore, we compare the complexity of STBC-CNOMA  with existing cooperative NOMA schemes in Section~\ref{sec:complexity_analysis}. In Section~\ref{sec:simulation_results}, we present numerical and simulation results, and conclusions are drawn in Section~\ref{sec:conclusions}.

\emph{Notation:} $\E[\cdot]$ and $\Var[\cdot]$ denote the statistical expectation and variance, respectively. In addition, $|\cdot|$ denotes the absolute value of a scalar quantity. Also, the definitions of the variables used in our analysis are provided in Table~\ref{Table1}.

\begin{table*}[t!] 
\caption{Table of Notations}
\label{Table1}
\centering
\begin{tabular}{P{1.1cm}|P{14cm}}
\hline
\cline{1-2}
Symbol & Definition\\
\hline\hline
$h_k$   & Channel gain from BS to the $k^{th}$ user \\
\hline
$g_{k,j}$   &Channel gain between the $k^{th}$ and the $j^{th}$ user \\
\hline
$\gamma_{k,noma}$   & SINR at the $k^{th}$ user to detect its own signal in direct NOMA phase \\
\hline
$\gamma_{k,ccn}$   &  SINR at the $k^{th}$ user for conventional cooperative NOMA case~\cite{ZDCNOMA} \\
\hline
$\gamma_{k}$   & SINR at the $k^{th}$ user with perfect timing synchronization, perfect SIC (pSIC), and perfect CSI (pCSI)\\
\hline
$\gamma_{k}^\eta$   & SINR at the $k^{th}$ user with perfect timing synchronization, imperfect SIC (ipSIC), and perfect CSI (pCSI)\\
\hline
$\gamma_{k}^\varepsilon$   & SINR at the $k^{th}$ user with imperfect timing synchronization, perfect SIC (pSIC), and perfect CSI (pCSI)\\
\hline
$\gamma_{k}^{\varepsilon,\eta}$   & SINR at the $k^{th}$ user with imperfect timing synchronization, imperfect SIC (ipSIC), and perfect CSI (pCSI)\\
\hline
\textcolor{black}{
$\gamma_{k}^{\chi}$}   & \textcolor{black}{SINR at the $k^{th}$ user with perfect timing synchronization, perfect SIC (pSIC), and imperfect CSI (ipCSI)} \\
\hline

$\gamma_{th}$   & SINR threshold \\
\hline
$\Upsilon$   & Rate threshold \\
\hline
$p_k $   & Power received at the $k^{th}$ user from BS  \\
\hline
$p_s$   & Mean power received by the user during STBC cooperation phase \\
\hline
$\xi_{k,t}$   &  Noise received at the $k^{th}$ user during time slot $t$ of the STBC cooperation phase\\
\hline
$r_{k,t}$   & Received signal at the $k^{th}$ user during time slot $t$ of the STBC cooperation phase   \\
\hline
$\lambda_h$   & Fading parameter for exponentially distributed variable $A$ \\
\hline
$\lambda_i$   &  Fading parameter for hypo-exponentially distributed variable $B$ \\
\hline
$\lambda_\eta$   & Fading parameter for exponentially distributed variable $F$  \\
\hline
$\lambda_g$   & Fading parameter for Gamma distributed variable $Z$  \\
\hline
$\erf(.)$ & Error function\\
\hline
$\mathbf{Ei}(x)$ & Exponential integral of $x$ and $\mathbf{Ei}(x)= \int_{-\infty}^{x} \frac{e^t}{t} dt$\\
\hline
\end{tabular}
\end{table*}

\section{System Model}
\label{sec:system_model}

\begin{figure}[t]
  \begin{center}
 \centerline{\includegraphics[scale=0.5]{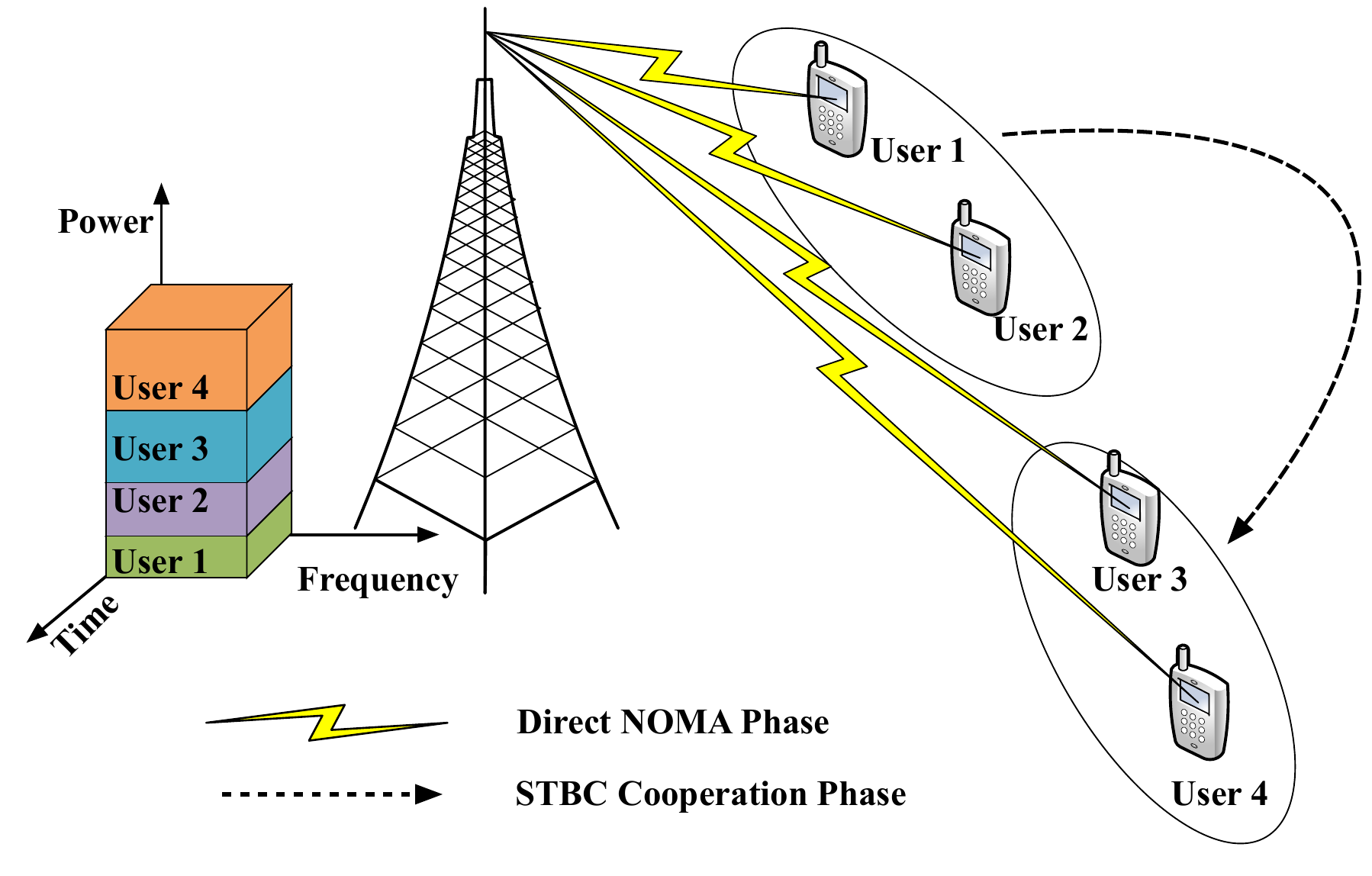}}
 \caption{An example illustration of downlink STBC-CNOMA with four users.}\label{stbc_noma_nw1}
 \end{center}
\end{figure}
We consider an STBC-based downlink NOMA system as shown in Fig.~\ref{stbc_noma_nw1}.  
 Base station~(BS) transmits the superimposed signal to all users in its coverage area. We assume the channel between the BS and the users and that between any two users to be flat fading Rayleigh channel, as in~\cite{ZDCNOMA} and~\cite{CRS-NOMA}. In general, the users near the BS experience a strong channel to the BS, henceforth referred to as the \textit{strong users}. Similarly, the users lying at the cell edge have weak channel conditions, and they are considered as \textit{weak users}. The user with the weakest channel conditions is assigned the maximum power, whereas the user with the strongest channel conditions is assigned the lowest power. Without loss of generality, it is assumed that the users are aligned as per descending order of their channel condition, i.e., $|h_1|\geq|h_2| \geq \dots \geq|h_k|\geq \dots \geq|h_K|$, where $|h_k|$ is the channel coefficient from BS to the $k^{th}$ user and $K$ is the total number of users. We consider User 1, $U_{1}$, as the strongest user and User $K$, $U_K$, as the weakest user, where $ \{U_1,U_2,\ldots, U_k,\ldots,U_K\}$ is the set of all users.

Transmission from BS to the users takes place in two phases. In the first phase, called the \textit{direct NOMA phase}, BS sends the superimposed signal to all users. The weakest user extracts its own signal by considering the signals for all the other users as noise. Other users employ SIC to cancel the interference from the weak users and treat the signals for other strong users as noise. In the second phase, referred to as \textit{cooperative NOMA phase}, \textcolor{black}{the first} two strongest users, $U_1$ and $U_{2}$, make an STBC pair and transmit the messages of next two users, $U_3$ and $U_{4}$, by a distributed $2\times2$  STBC transmission. This process of $2\times2$ STBC  continues until the weakest user $U_K$ is reached. 

\subsection {Direct NOMA Phase}
\begin{figure}[t]
  \begin{center}
 \centerline{\includegraphics[scale=0.4]{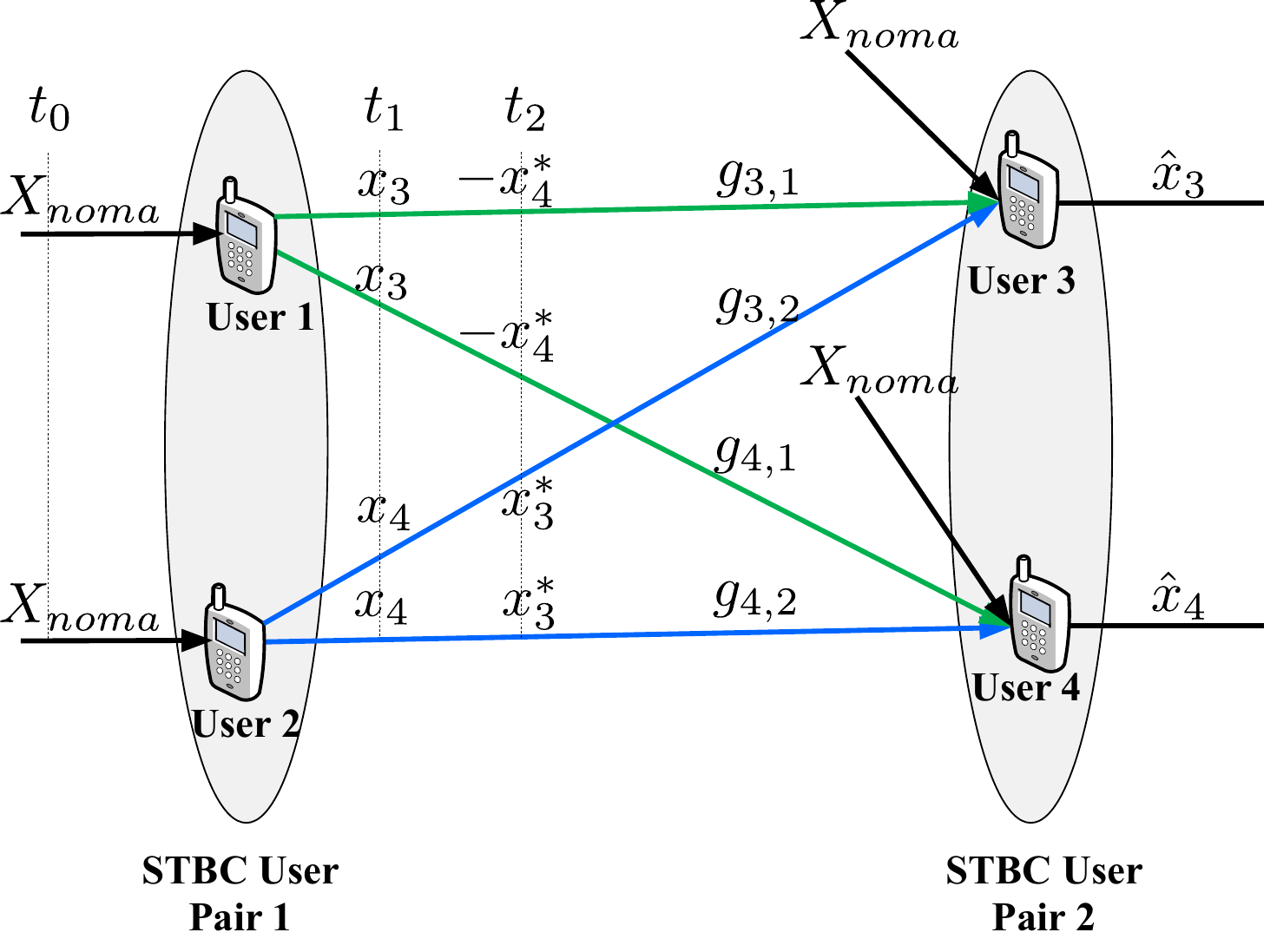}}
 \caption{Cooperation mechanism in the STBC-CNOMA network with two STBC user pairs.}\label{stbc_noma_nw2}
  \vspace{-15pt}
 \end{center}
\end{figure}

As shown in Fig.~\ref{stbc_noma_nw2}, the direct NOMA phase is accomplished in the first time slot, when the BS transmits the superimposed signal to all of the $K$ users.
The $k^{th}$ user, such that $ 1 \leq k < K$, detects the message of \textcolor{black}{the $i^{th}$} user, where \textcolor{black}{$i>k$}, then applies SIC to subtract it from the superimposed signal and finally detects its own message. 

 \subsection{STBC-based Cooperative Transmission Phase}
The second phase of the proposed transmission is the STBC-based cooperative transmission phase. \textcolor{black}{Users are paired as per their channel conditions, i.e., the first two strongest users make the first user pair followed by $U_3$ and $U_4$ making the second user pair until $U_{K-1}$ and $U_{K}$ making the $M^{th}$ user pair, where $M=K/2$ and $K$ is even. For the case $K$ is odd, $U_{K-2}$ and $U_{K-1}$ construct the $M^{th}$ user pair, where $M=\frac{K-1}{2}$.} In this phase, all of the users cooperate with each other by employing a distributed $2 \times 2$ STBC transmission. However, at the receiving STBC users' pair, we use  $2 \times 1$  STBC reception for the detection of the symbols~\cite{Alamouti}. Each receiving STBC user receives two symbols, one for itself and the other for its neighbor. Thus, a STBC user keeps the decoded symbol for itself, while the other symbol for its neighbor is discarded.  
In the first time slot, $t_0$, BS transmits the composite NOMA signal to all users in its coverage area. Since \textcolor{black}{the first} two strongest users have decoded the messages for all the other users, therefore, they can contribute in the STBC cooperation by transmitting the information for next two users in the next two time slots. Therefore, during the second and third time slots, $t_1$ and $t_2$, $U_1$ and $U_{2}$ transmit to the users $U_3$ and $U_{4}$ using Alamouti code. Similarly, in the two following time slots, $U_3$ and $U_{4}$ transmit the STBC signal to $U_{5}$ and $U_{6}$ and this process continues till $U_K$ receives its message.
Assuming 
 The SINR at each user of the receiving STBC pair, with perfect timing synchronization and perfect SIC, is given by
\begin{equation}\label{sinr_k_STBC-CNOMA}
\gamma_{k} =  \frac{|h_{k}|^2 p_{k} }{\sum\limits_{i=1}^{k-1}{|h_{k}|^2 p_{i} + \sigma^2}} + \frac{(|g_{k,k-\iota-1}|^2+|g_{k,k-\iota-2}|^2) {p}_{s} }{ \sigma^2},\\{}
\end{equation}
where $2<k\leq K$, $p_{k}=\Phi_k P_{NOMA}$ is the power assigned to the $k^{th}$ user, $\Phi_k$ is the power coefficient for the $k^{th}$ user, $P_{NOMA}$ is the power assigned to the composite NOMA signal and ${p}_{s}$ is the fraction of power transmitted from the transmitting users' pair in STBC cooperation. Also, $\iota\in \lbrace 0,1 \rbrace$ denotes \textcolor{black}{the first and the second} user of the $2 \times 2$ STBC receiving pair, respectively. In case of conventional cooperative NOMA~\cite{ZDCNOMA}, each strong user, $U_i$, for any $i<k$ will cooperate with the weak user, $U_k$, by means of decode and forward relay. \textcolor{black}{Assuming maximum-ratio combining~(MRC), as in~\cite{ZDCNOMA}, the SINR at each user in this case is given by 
\begin{equation}\label{sinr_k_CNOMA}
\begin{aligned}
\gamma_{{k}_{ccn}} = {} & \frac{|h_{k}|^2 p_{k} }{\sum\limits_{i=1}^{k-1}{|h_{k}|^2 p_{i} + \sigma^2}} +\sum\limits_{j=1}^{k-1}\frac{|g_{k,k-j}|^2 {q}_{k,k-j} }{ \sum\limits_{i=1}^{k-1}{|g_{k,k-j}|^2 {q}_{k,k-i} + \sigma^2}},
\end{aligned}
\end{equation}
 where $\gamma_{{k}_{ccn}}$ is the SINR at the $k^{th}$ user for conventional cooperative NOMA and ${q}_{k,k-j}$ is the power transmitted from the $(k-j)^{th}$ user to the $k^{th}$ user in the cooperation phase. }

\textcolor{black}{
\section{Three Practical Impairments}
\label{sec:three_practical_impairments}
In this section, we analyze the STBC-CNOMA system in the presence of the timing mismatch (or synchronization error) in the STBC cooperation phase, imperfect SIC, and channel estimation error. We treat the three imperfections in the three subsections, separately. To better explain the impacts of the three practical impairments, we use a simple example with four users (i.e., $K=4$) as shown in  Fig.~\ref{stbc_noma_nw2}. Using this example, we analyze the STBC transmission from $U_1$ and $U_{2}$ to $U_3$ and $U_{4}$. Then, based on the same approach, we extend our analysis to a general scenario with more number of users (i.e., $K>4$).
}
%
%
%
%
%
%

\subsection{Synchronization Error}
Fig.~\ref{stbc_noma_nw2} shows an STBC-based downlink NOMA network for two user pairs. During \textcolor{black}{the first} time slot $t_o$, each user receives the composite NOMA signal $X_{noma}$ from the base station. Since $U_1$ and $U_{2}$ are located in close vicinity of the BS, they decode their own messages in addition to the messages of $U_3$ and $U_4$ and send these to $U_3$ and $U_{4}$ through STBC transmission. In other words, $U_1$ and $U_{2}$ send $x_3$ and $x_4$ to $U_3$ and $U_{4}$ during time slot $t_1$. During next time slot $t_2$, $U_1$ and $U_{2}$ send $-x_{4}^*$  and $x_{3}^*$, respectively, to $U_3$ and $U_{4}$. Thus, the STBC receiving user pair $U_3$ and $U_4$ can detect their respective messages.

\begin{figure}[t!]
\begin{subfigure}{.5\textwidth}
  \centering
  \includegraphics[scale=0.3]{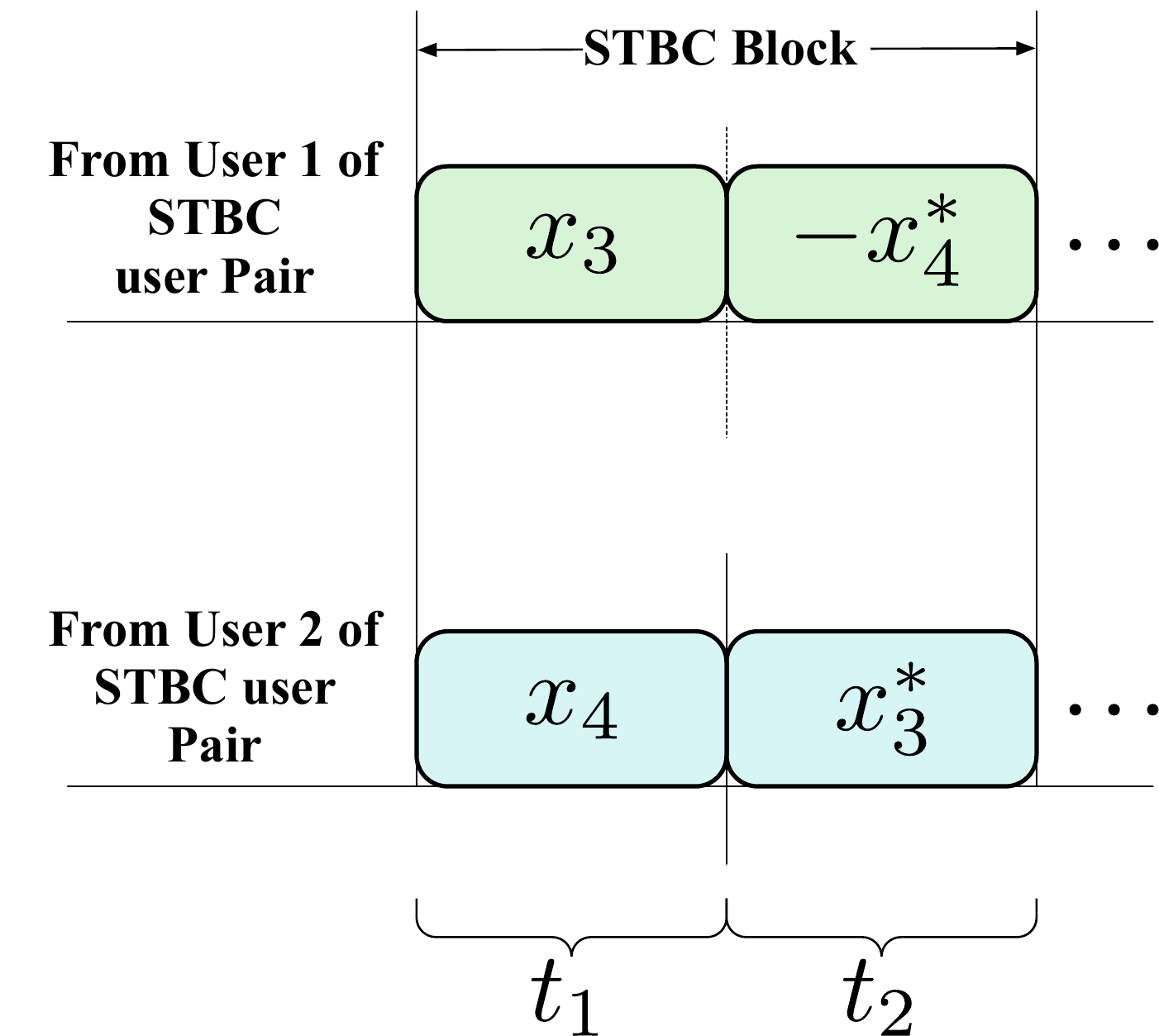}  
  \caption{Perfect timing synchronization}
  \label{fig:sub-first}
  \vspace{7pt}
\end{subfigure}
\begin{subfigure}{.5\textwidth}
  \centering
  \includegraphics[scale=0.3]{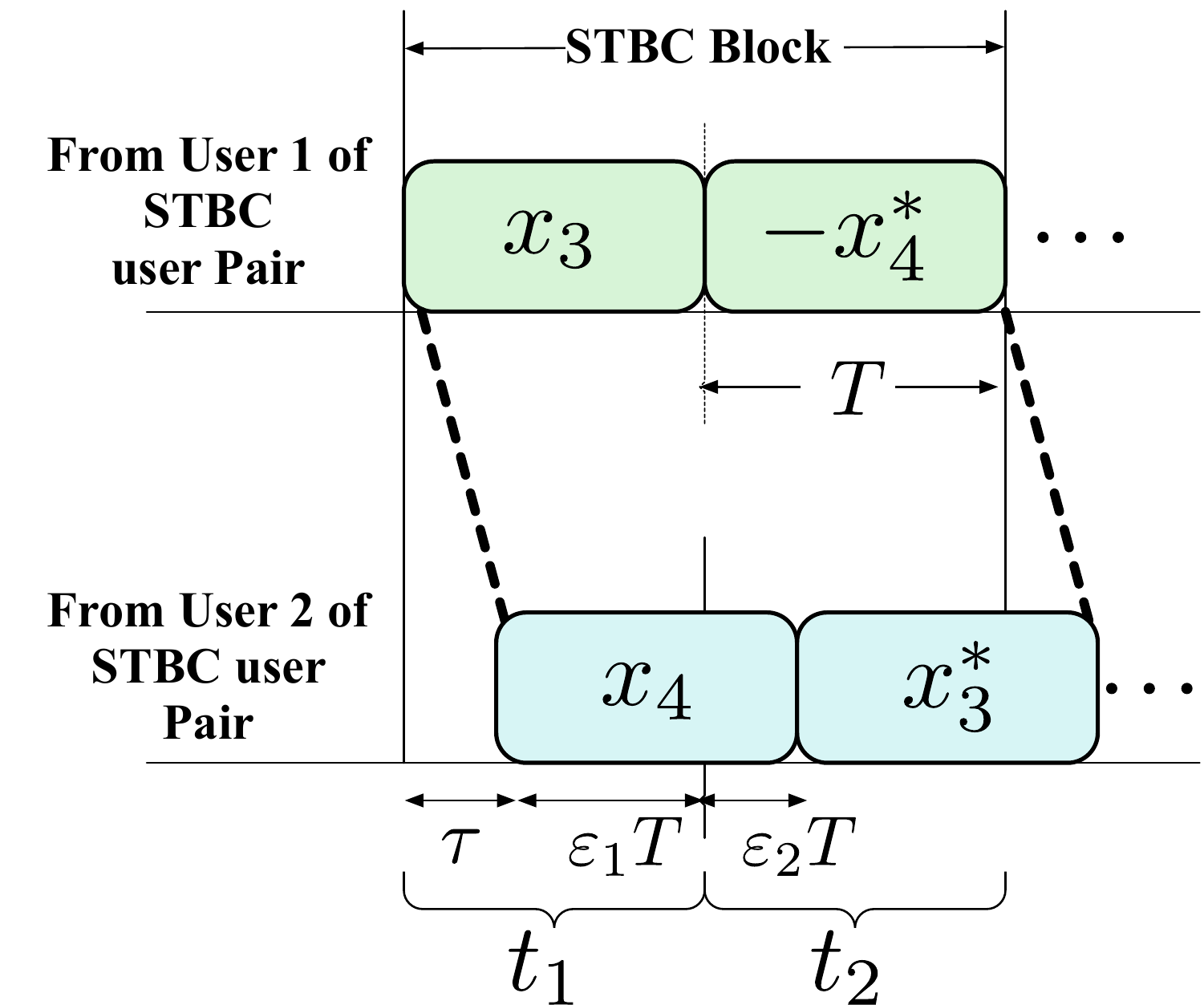}  
  \caption{Imperfect timing synchronization with $\uptau = \varepsilon_2T$}
  \label{fig:sub-second}
\end{subfigure}
\caption{Timing diagram of the received signals.}
\label{ISI0}
\end{figure}

Fig.~\ref{ISI0} illustrates the timing diagram with different synchronization conditions. Fig.~\ref{ISI0}(a) depicts the STBC mechanism at the receiver with perfect timing synchronization. The symbols from both users $U_1$ and $U_2$ of STBC pair arrive at the receiver at same time instant, where $T$ is the symbol duration. 
On the other hand, Fig.~\ref{ISI0}(b) shows the STBC transmission with timing offset of $\uptau =  \varepsilon_2 T$. In this case, the symbols from $U_1$ and $U_2$ does not arrive simultaneously and there is a substantial inter-symbol-interference (ISI) experienced by the STBC receiving user pair, which causes the decrease in the SINR. The STBC block used by $U_1$ and $U_{2}$ is given by 
\begin{equation} \label{mat1}
S=
  \begin{bmatrix}
    x_{3} &  -x_{4}^*\\
    x_{4} &  x_{3}^*
  \end{bmatrix}.
\end{equation}
Thus, if $U_{4}$ is not perfectly synchronized, as shown in Fig.~\ref{ISI0}(b), the receiver equations for $U_{3}$ or $U_{4}$ of STBC receiving pair at time slots $t_1$ and $t_2$ are given as
\begin{align}
r_{3,1} &= g_{3,1}x_{3} + g_{3,2}\varepsilon_1x_{4} + \xi_{3,1}, \\
r_{3,2} & = -g_{3,1}x_{4}^* + g_{3,2}\varepsilon_1x_{3}^* +g_{3,2}\varepsilon_2x_{4} +\xi_{3,2},\\
r_{4,1} &= g_{4,1}x_{3} + g_{4,2}\varepsilon_1x_{4} + \xi_{4,1}, \\
r_{4,2} & = -g_{4,1}x_{4}^* + g_{4,2}\varepsilon_1x_{3}^* +g_{4,2}\varepsilon_2x_{4} +\xi_{4,2},
\label{r1_ep=notzero}
\end{align}
where $r_{k,t}$ and $\xi_{k,t}$ are the received signal and the additive noise observed at the $k^{th}$ user during time slot $t$, respectively. The received signals after combiner at User 3 and User 4 are given by 
\begin{align}\label{v23a}
\tilde{v}_{3} &= g_{3,1}^*r_{3,1} + g_{3,2}r_{3,2}^*, \\
\tilde{v}_{4} & = g_{4,2}^*r_{4,1} - g_{4,1}r_{4,2}^*,
\label{v23b}
\end{align}
which can be expanded as
\begin{align}
\tilde{v}_{3} & = (|g_{3,1}|^2 + \varepsilon_1|g_{3,2}|^2)x_{3} +(\varepsilon_1 - 1)g_{3,1}^*g_{3,2}x_{4} \nonumber\\
{}&+\varepsilon_2|g_{3,2}|^2x_{4}^*+ g_{3,1}^*\xi_{3,1} + g_{3,2}\xi_{3,2}^*,\\
  \tilde{v}_{4} &= (|g_{4,1}|^2 + \varepsilon_1|g_{4,2}|^2)x_{4} +(1 - \varepsilon_1)g_{4,1}g_{4,2}^*x_{3}\nonumber\\
  {}&-\varepsilon_2g_{4,1}g_{4,2}^*x_{4}^*+ g_{4,2}^*\xi_{4,1} - g_{4,1}\xi_{4,2}^*.
\label{y12}
\end{align}

\begin{figure*}[h!]
\normalsize
\begin{equation}
\gamma_{3}^{\varepsilon} =
 \frac{|h_{3}|^2 p_{3} }{\sum\limits_{i=1}^{2}{|h_{3}|^2 p_{i} + \sigma^2}} +\frac{( |g_{3,1}|^2 + \varepsilon_1|g_{3,2}|^2)^2p_{s}}{|(\varepsilon_1 - 1)g_{3,1}^*g_{3,2}+  \varepsilon_2g_{3,2}g_{3,2}^*|^2 p_{s}+{(|g_{3,1}|^2 + |g_{3,2}|^2)\sigma^2}}.
\label{SINR_3_ep}
\end{equation}
\begin{equation}
\gamma_{4}^{\varepsilon} = 
 \frac{|h_{4}|^2 p_{4} }{\sum\limits_{i=1}^{3}{|h_{4}|^2 p_{i} + \sigma^2}} +\frac{ (|g_{4,1}|^2 + \varepsilon_1|g_{4,2}|^2)^2p_{s}}{{|(1-\varepsilon_1 )g_{4,1}g_{4,2}^*}-{\varepsilon_2g_{4,1}g_{4,2}^*|^2 p_{s}}+{(|g_{4,1}|^2 + |g_{4,2}|^2)\sigma^2}}.
\label{SINR_4_ep}
\end{equation}
\begin{align}
\gamma_{m}^{\varepsilon} &= 
 \frac{|h_{m}|^2 p_{m} }{\sum\limits_{i=1}^{m-1}{|h_{m}|^2 p_{i} + \sigma^2}} +\frac{ (|g_{m,m-2}|^2 + \varepsilon_1|g_{m,m-1}|^2)^2p_{s}}{{|(\varepsilon_1 - 1)g_{m,m-2}^*g_{m,m-1}+  \varepsilon_2g_{m,m-2}g_{m,m-2}^*|^2p_{s}}+{(|g_{m,m-2}|^2 + |g_{m,m-1}|^2)\sigma^2}}.
 \label{SINR_k_odd_ep}
\end{align}
\begin{align}
\gamma_{n}^{\varepsilon} &=\frac{|h_{n}|^2 p_{n} }{\sum\limits_{i=1}^{n-1}{|h_{n}|^2 p_{i} + \sigma^2}} +\frac{ (|g_{n,n-3}|^2 + \varepsilon_1|g_{n,n-2}|^2)^2p_{s}}{{|(1-\varepsilon_1 )g_{n,n-3}g_{n,n-2}^*}-{\varepsilon_2g_{n,n-3}g_{n,n-2}^*|^2p_{s}}+{(|g_{n,n-3}|^2 + |g_{n,n-2}|^2)\sigma^2}}.
 \label{SINR_k_even_ep}
 \end{align}
\begin{align}
\gamma_{m}^{\varepsilon,\eta} &= 
 \frac{|h_{m}|^2 p_{m} }{\eta  |g_{\eta}|^2 p_{\eta}  + \sum\limits_{i=1}^{m-1}{|h_{m}|^2 p_{i} + \sigma^2}} +\frac{ (|g_{m,m-2}|^2 + \varepsilon_1|g_{m,m-1}|^2)^2p_{s}}{{|(\varepsilon_1 - 1)g_{m,m-2}^*g_{m,m-1}+  \varepsilon_2g_{m,m-2}g_{m,m-2}^*|^2p_{s}}+{(|g_{m,m-2}|^2 + |g_{m,m-1}|^2)\sigma^2}}.
 \label{SINR_k_odd_ep_eta}\\
\gamma_{n}^{\varepsilon,\eta} &=\frac{|h_{n}|^2 p_{n} }{\eta  |g_{\eta}|^2 p_{\eta} +\sum\limits_{i=1}^{n-1}{|h_{n}|^2 p_{i} + \sigma^2}} +\frac{ (|g_{n,n-3}|^2 
+ \varepsilon_1|g_{n,n-2}|^2)^2p_{s}}{{|(1-\varepsilon_1 )g_{n,n-3}g_{n,n-2}^*}-{\varepsilon_2g_{n,n-3}g_{n,n-2}^*|^2p_{s}}+{(|g_{n,n-3}|^2 + |g_{n,n-2}|^2)\sigma^2}}.
 \label{SINR_k_even_ep_eta}
\end{align}
\textcolor{black}{
\begin{align}
\gamma_{m}^{\chi} &= 
 \frac{|h_{m}|^2 p_{m} }{\sum\limits_{i=1}^{m-1}{|h_{m}|^2 p_{i} + \sigma^2}} +\frac{(|\varrho_{m,m-2}+g_{m,m-2} \rho|^2 + |\varrho_{m,m-1}+g_{m,m-1} \rho|^2)p_{s}}{\sigma^2}.
 \label{SINR_k_odd_chi}\\
\gamma_{n}^{\chi} &=\frac{|h_{n}|^2 p_{n} }{\sum\limits_{i=1}^{n-1}{|h_{n}|^2 p_{i} + \sigma^2}}  +\frac{(|\varrho_{n,n-3}+g_{n,n-3} \rho|^2 + |\varrho_{n,n-2}+g_{n,n-2} \rho|^2)p_{s}}{\sigma^2}.
 \label{SINR_k_even_chi}
\end{align}
}
\hrulefill
\end{figure*}
%
As a result, assuming MRC of the received signals in both direct NOMA and STBC phases, the SINRs of $U_3$ and $U_{4}$ can be obtained as~\eqref{SINR_3_ep} and \eqref{SINR_4_ep}, respectively. We note that if timing synchronization is perfect (i.e., $\varepsilon_1 = 1$ and $\varepsilon_2 = 0$), the corresponding SINRs in \eqref{SINR_3_ep} and \eqref{SINR_4_ep} are reduced into \eqref{sinr_k_STBC-CNOMA}.
Furthermore, generalizing this four-user example to a larger number of users (e.g., $K=6, 8, ...$), in the presence of the synchronization error, the mathematical expressions of the SINR of the $m^{th}$ and the $n^{th}$ users can be derived as \eqref{SINR_k_odd_ep} and \eqref{SINR_k_even_ep}, respectively, for any $m\in\{3,5,7,\dots,K-1\}$ and $n\in\{4,6,8,\dots,K\}$.

\subsection{Imperfect SIC}
Without synchronization error but under the residual interference caused by imperfect SIC implementation, 
the SINRs at the $m^{th}$ and the $n^{th}$ users are given by
\begin{align}
\gamma_{m}^\eta &= {} 
 \frac{|h_{m}|^2 p_{m} }{\eta  |g_{\eta}|^2 p_{\eta}+\sum\limits_{i=1}^{m-1}{|h_{m}|^2 p_{i} + \sigma^2}} \nonumber\\
 {}&+\frac{ (|g_{m,m-1}|^2 + |g_{m,m-2}|^2)p_{s}}{\sigma^2},
 \label{SINR_m_pTim_ipSIC}
\end{align}
 and 
\begin{align}
\gamma_{n}^\eta &= {} 
 \frac{|h_{n}|^2 p_{n} }{\eta  |g_{\eta}|^2 p_{\eta}+\sum\limits_{i=1}^{n-1}{|h_{n}|^2 p_{i} + \sigma^2}} \nonumber\\
 {}&+\frac{ (|g_{n,n-2}|^2 + |g_{n,n-3}|^2)p_{s}}{\sigma^2},
 \label{SINR_n_pTim_ipSIC}
\end{align}
respectively.
Therefore, in the presence of both the synchronization error and imperfect SIC,
the SINRs of the $m^{th}$ and the $n^{th}$ users can be derived as 
%
\eqref{SINR_k_odd_ep_eta} and ~\eqref{SINR_k_even_ep_eta}, respectively, where $m\geq3$ and $n\geq4$ can be any odd and even numbers, respectively. In addition, it is to be noted that $\eta=0$ and $\eta=1$ represent the perfect and imperfect SIC employed at that user, respectively.  \textcolor{black}{It is noted that the SIC imperfection and timing error do not affect each other.}


\subsection{Imperfect CSI}
\textcolor{black}{In this section, we consider the imperfect CSI (i.e., channel estimation error) along-with imperfect timing synchronization. We consider $\hat{g}_{k,j}= g_{k,j} + \omega_j$, where $\hat{g}_{k,j}$ is the estimate of $g_{k,j}$ and $g_{k,j}$ is the channel between $k^{th}$ and $j^{th}$ user. $\omega_j $ is the channel estimation error and it is assumed to be complex Gaussian random variable (RV) with zero mean and variance of $\sigma_\omega^2$. The RV $\hat{g}_{k,j}$ are complex Gaussian with zero mean and variance $\sigma^2_{\hat{g}}= \sigma^2_{g}+\sigma^2_\omega$. Also, the correlation coefficient between the estimated channel and the real channel is  $\rho=\sigma_g^2/(\sigma_g^2+\sigma_{\omega}^2)$. We can write that $g_{k,j} = \rho \hat{g}_{k,j} + \varrho_{k,j}$, where $\varrho_{k,j}$ are independent complex Gaussian RVs with zero mean and variance $\sigma_ {\varrho}^2= \sigma_{g}^2\sigma_\omega^2/(\sigma_{g}^2+\sigma_\omega^2)$~\cite{ipCSI1,ipCSI2}. By putting the value  $ g_{k,j} = \rho \hat{ g_{k,j} } + \varrho_{k,j}$ into~\eqref{v23a} and ~\eqref{v23b}~\cite{stbc1}, we get }
\textcolor{black}{
\begin{align}
\tilde{v}_{3}^\chi &={}
[ \left(\varrho_{3,1}+g_{3,1} \rho \right) \left(\varrho_{3,1}+\rho 
   g_{3,1}\right){}^*\nonumber\\
  {}&+ \left(\varrho_{3,2}+g_{3,2} \rho \right) \left(\varrho_{3,2}+\rho  g_{3,2}\right){}^*]x_3\nonumber\\
  {}&+ \xi_{3,1} \left(\varrho_{3,1}+\rho  g_{3,1}\right){}^*+\xi_{3,2}^* \left(\varrho_{3,2}+g_{3,2} \rho \right),\label{24}
\end{align} 
}
\textcolor{black}{
and
\begin{align}
\tilde{v}_{4}^\chi &={}
   [\left(\varrho_{4,2}+g_{4,2} \rho \right) \left(\varrho_{4,2}+\rho 
   g_{4,2}\right){}^*\nonumber\\
  {}&+ \left(\varrho_{4,1}+g_{4,1} \rho \right) \left(\varrho_{4,1}+\rho  g_{4,1}\right){}^*]x_4 \nonumber\\
  {}&\xi_{4,1} \left(\varrho_{4,2}+\rho  g_{4,2}\right){}^*-\xi_{4,2}^* \left(\varrho_{4,1}+g_{4,1} \rho \right),\label{25}
 \end{align}
 respectively.}
\textcolor{black}{
 Assuming the MRC of the received signals at each user for direct NOMA and STBC cooperation phase, and solving \eqref{24}-\eqref{25} for the SINRs, we obtain \eqref{SINR_k_odd_chi} and \eqref{SINR_k_even_chi}, respectively.
}

\section{Outage Probability Analysis}
\label{sec:outage_probability_analysis}
\textcolor{black}{In this section, we analyze the outage performance under the three practical impairments: timing error, imperfect SIC, and channel estimation error.} Because the last user (i.e., User $K$) has the weakest channel gain and also suffers from the impairments, it has the worst outage probability compared to the other users, as shown in~\cite{jamal2017new, jamal2018efficient}. For this reason, we focus on the outage performance of User $K$ (e.g., User 4 in the four-user example in the previous section), which will set a benchmark for the other users with stronger channel gains.

An outage event occurs when a user cannot achieve the reliable SINR to detect the signal. Following~\cite{jamal2018efficient} and~\cite{jamal2017new},  the outage probability of any user $k$ (for $k\leq K$) is defined as
\textcolor{black}{
\begin{equation}
\begin{aligned}
 P_{out} = {} &\mathop{\mathbb{P}}(\gamma_{k} < \gamma_{th})=\int_{0}^{\gamma_{th}} f_\Gamma(\gamma_{k}) d\gamma_{k}, \label{Pout_k}
\end{aligned}
\end{equation}
}
where $\gamma_{th}$ is the SINR threshold and \textcolor{black}{$f_\Gamma(\gamma_{k})$ is the probability density function~(PDF) of SINR received at the $k^{th}$ user with perfect SIC, perfect timing synchronization, and perfect CSI, which is derived in \eqref{sinr_k_STBC-CNOMA}. We can also use the SINR derived in the previous section as in \eqref{SINR_k_odd_ep}-\eqref{SINR_k_even_chi} for different cases in order to find their respective outage probabilities.} Similarly, the rate outage (i.e., capacity outage) probability is defined as
\begin{equation}
\textcolor{black}{
 \tilde{P}_{out}  =  \mathop{\mathbb{P}}[\gamma_{k} < 2^\Upsilon-1],
 }
\label{Pout_k1}
\end{equation}
where $\Upsilon=\log_2(1+\gamma_{th})$ is the rate threshold. 
In order to find the outage probability using \eqref{Pout_k}, we need the PDF of SINR for different cases. Therefore, as the SINR expressions in \eqref{SINR_k_odd_ep}-\eqref{SINR_k_even_chi} contain random variables, we consider the following mathematical manipulation and define some composite random variables for finding the PDF of SINRs for different cases. 

We redefine the variables used in \eqref{SINR_k_odd_ep}-\eqref{SINR_k_even_chi}  as $A=|h_k|^2p_k$, $B=\sum\limits_{i=1}^{I}|h_k|^2p_i$, $C=|g_{k,k-2}|^2 p_s$, $D=|g_{k,k-3}|^2p_s$, $F=|g_{\eta}|^2 p_{\eta}$, $C_{\chi}=|A_{\chi}|^2 p_{s}$ and $D_{\chi}=|B_{\chi}|^2 p_{s}$, where $|h_k|^2$, $|g_{\eta}|^2$, $|g_{k,k-2}|^2$, $|g_{k,k-3}|^2$, $|A_{\chi}|^2$ and $|B_{\chi}|^2$ follow the exponential distribution with parameters $\zeta_h$, $\zeta_\eta$, $\zeta_{g_{k,k-2}}$, $\zeta_{g_{k,k-3}}$, $\zeta_{a_\chi}$ and $\zeta_{b_\chi}$, respectively. The variables $A$,  $C$, $D$, $F$, $C_{\chi}$ and $D_{\chi}$ also follow the exponential distributions with parameters $\lambda_h$, $\lambda_{g_{k,k-2}}$, $\lambda_{g_{k,k-3}}$, $\lambda_\eta$, $\lambda_{a_\chi}$ and $\lambda_{b_\chi}$, respectively. \textcolor{black}{We assume that $\lambda_{g_{k,k-2}}$ = $\lambda_{g_{k,k-3}}$ = $\lambda_{g}$ and $\lambda_{a_\chi}=\lambda_{b_\chi}=\lambda_{\chi}$, where disparate path losses with large-scale fading are compensated by appropriate power control at the relaying users, as in~\cite{DSTBC_Uysal2007}.} 
The variable $B$ follows the hypo-exponential distribution with parameters $\lambda_i$, where $i\in(1,2,3,\dots,I)$ is a set of interfering users and $I=k-1$. We denote $\lambda_h=\frac{1}{p_k\zeta_h}$, $\lambda_i=\frac{1}{p_i \zeta_i}$,  $\lambda_g=\frac{1}{p_s\zeta_g}$, $\lambda_\eta=\frac{1}{p_\eta\zeta_\eta}$ 
\textcolor{black}{
and  $\lambda_\chi=\frac{1}{p_\chi\zeta_\chi}$, where $p_\eta$ and $p_\chi$ is the power of interfering signal~(IS) due to imperfect SIC and power of IS due to imperfect CSI, respectively.}


\textcolor{black}{
In the following five lemmas and five propositions, we treat different combinations of the three impairments. In the lemmas, we derive the exact outage probabilities based on  $\tilde{P}_{out}$ in \eqref{Pout_k1}. The probability distributions of the SINRs in each case can be found in its proof. Further, in the propositions, which correspond to each of the five lemmas, we provide the asymptotic outage probabilities in the high transmit SNR regime using $\tilde{P}_{out}$ in \eqref{Pout_k1}, which provide intuitive insights into the impacts of the impairments. 
}
For mathematical notations used to derive the lemmas, please refer to Table \ref{TableII}.

\begin{table}[t] 
\caption{Mathematical Notations}
\label{TableII}
\begin{tabular}{|P{8cm}|}
\hline
\cline{1-1}
Mathematical Notations used in Corollaries and Appendices \\
\hline\hline
$\psi_1=\displaystyle\prod_{j=1}^{I} \lambda_j$, $\psi^{\eta}_1 =\lambda_h \lambda_{\eta}\displaystyle\prod_{j=1}^{I} \lambda_j$, $\psi_{2}=\sum\limits_{i=1}^{I} \displaystyle\prod_{\mathclap{\substack{j=1,\\j\neq i}}}^{I} \lambda_j$,    \\
\hline
$\psi^{\eta}_{2}=\lambda_{\eta}+\psi_2$, $\psi_{3}=\sum\limits_{i=1}^{I} log[\lambda_i] \displaystyle\prod_{\mathclap{\substack{j=1,k=1,\\j\neq i,j\neq k,\\i\neq k}}}^{I} (\lambda_j-\lambda_k)$, $\psi_{4}=\lambda_h \psi_{a}$,  $\psi_{3}^\eta=\sum\limits_{j=1}^{I}log[\lambda_i] \displaystyle\prod_{\mathclap{\substack{j=1,k=1,\\i\neq j,i\neq k,\\j\neq k}}}^{I} (\lambda_j-\lambda_k) +log[\lambda_\eta] \displaystyle\prod_{\mathclap{\substack{j=1,k=1,\\i\neq j,i\neq k,\\j\neq k}}}^{I} (\lambda_j-\lambda_k)$ \\
\hline
  $\psi_{4}^\eta=\lambda_h\sum\limits_{j=1}^{I}log[\lambda_i] \displaystyle\prod_{j=1,k=1,k>j}^{I} (\lambda_j-\lambda_k)(\lambda_j-\lambda_\eta)$, $\psi_{6}=\sum\limits_{i=1}^{I} log[\lambda_i] \displaystyle\prod_{\mathclap{\substack{j=1,k=1,\\j\neq i, i\neq k,\\k>j}}}^{I} (\lambda_j-\lambda_k)$, \\
\hline
 $\psi_{7}=\displaystyle\prod_{\mathclap{\substack{j=1,k=1,\\k>j}}}^{I} (\lambda_j-\lambda_k)^2$, $\psi_{8}=\sum\limits_{i=1}^{I}\lambda_i log[\lambda_i] \displaystyle\prod_{\mathclap{\substack{j=1,k=1,\\j\neq i,i\neq k,\\k>j}}}^{I} (\lambda_j-\lambda_k)$,\\
\hline $\psi_{7}^\eta=\sum\limits_{i=1}^{I}log[\lambda_i] \displaystyle\prod_{\mathclap{\substack{j=1,k=1,\\i\neq j,i\neq k,\\i\neq \eta, k>j}}}^{I} (\lambda_j-\lambda_k)(\lambda_j-\lambda_\eta)
+log[\lambda_\eta] \displaystyle\prod_{\mathclap{\substack{j=1,k=1,\\i\neq j,i\neq k,\\i\neq \eta, k>j}}}^{I} (\lambda_j-\lambda_k)$,\\
\hline
 $\psi_{8}^\eta= \psi_{8} +\lambda_\eta log(\lambda_\eta)\displaystyle\prod_{\mathclap{\substack{j=1,j\neq \eta}}}^{I}(\lambda_j-\lambda_\eta)$, $\psi_{9}=\lambda_h^2 \psi_7$, \\
 \hline
$\psi_{9}^\eta=\lambda_h^2 \psi_{b} \displaystyle\prod_{\mathclap{\substack{j=1,\eta=1,\\\eta>j}}}^{I}(\lambda_j-\lambda_\eta)^2$, $\psi_{10}=\sum\limits_{i=1}^{I}e^{\lambda_i} \displaystyle\prod_{\mathclap{\substack{j=1,k=1,\\j\neq i,i\neq k,\\k>j}}}^{I} (\lambda_j-\lambda_k)E_i(-\lambda_i)$, $\psi_{10}^\eta=\lambda_h^2 \psi_{8}^\eta $\\
\hline
$\psi^{I}=\sum\limits_{i=1}^{I}\sum\limits_{j>i}^{I}\dots\sum\limits_{s>\dots>j>i}^{I} \lambda_s\dots\lambda_j\lambda_i$, $\psi_{a}=\displaystyle\prod_{\mathclap{\substack{j=1,k=1,\\k>j}}}^{I} (\lambda_j-\lambda_k)$,$\psi_{b}=\displaystyle\prod_{\mathclap{\substack{j=1,k=1,\\k>j}}}^{I} (\lambda_j-\lambda_k)^2$  \\
\hline
\end{tabular}
\end{table}


\textcolor{black}{First, we consider the outage probability in the absence of any impairments, which can serve as a baseline to quantify the impact fo the imperfections, in the following lemma.}
\begin{lemma}
The outage probability for the perfect timing, perfect SIC, and perfect CSI is given as
\end{lemma}
\begin{align}\label{out_gamma_22k}
P_{out} = {} & \frac{1}{\lambda_g^2 \lambda_h^2 }\sum\limits_{i=1}^{I}\Bigg[\textcolor{black}{\mho} e^{-\frac{\lambda_i+\lambda_h \gamma_{th}}{\lambda_g \lambda_h}} \Bigg(\lambda_i (\lambda_i+\lambda_h \gamma_{th}) \text{Ei}\left(\frac{\lambda_i}{\lambda_g \lambda_h}\right)\nonumber\\
&-\lambda_i (\lambda_i+\lambda_h \gamma_{th}) \text{Ei}\left(\frac{\gamma_{th} \lambda_h+\lambda_i}{\lambda_g \lambda_h}\right)+\lambda_g \lambda_h e^{\frac{\lambda_i}{\lambda_g \lambda_h}}\nonumber\\
&\bigg(\left(e^{\gamma_{th}/\lambda_g}-1\right) (\lambda_g \lambda_h+\lambda_i)-\lambda_h \gamma_{th}\bigg)\Bigg)\Bigg],
\end{align}
where $\gamma_{th}$ is the SINR threshold and \textcolor{black}{ $\mho=\sum\limits_{i=1}^{I}\frac{ \prod_{i=1}^{I} \lambda_i}{ \prod_{j=1,i\neq j}^{I}(\lambda_j-\lambda_i)}$.}
\subsubsection*{Proof } \textit{See Appendix~A.}\\
\textcolor{black}{Since the exact outage expression derived in Lemma 1 is complicated and does not yield to easy interpretation, we consider the asymptotic behavior of the rate outage given in~\eqref{Pout_k1}, when the transmit SNR, which is denoted by $\SNR$, is high enough in the following proposition.}

\begin{proposition}
\textcolor{black}{As $\SNR\rightarrow \infty$, the rate outage probability for the perfect timing synchronization, perfect SIC, and perfect CSI becomes}
\textcolor{black}{
\begin{align}
 \lim_{\SNR \to \infty}\tilde{P}_{out}   \sim  \frac{\lambda_h g(\SNR)}{\Phi_k-\sum\limits_{i=1}^{k-1}\Phi_i( 2^{\Upsilon}-1)}+ 2\lambda_{g}^2[g(\SNR)]^2,\label{AsPout_k2f}
\end{align}
where $g(\SNR)=\frac{2^{\Upsilon}-1}{\SNR}$.}
\end{proposition}

\subsubsection*{Proof } \textit{See Appendix~B.}\\
\textcolor{black}{
Based on this ideal case, we will investigate how each impairment impacts the outage probability.}

\begin{lemma}
The outage probability for the perfect timing synchronization, imperfect SIC, and perfect CSI is given as
\end{lemma}
\begin{align}
P_{out}^\eta &={} \sum\limits_{i=1}^{I}\Bigg[\frac{1}{\lambda_g^3 \lambda_h^2 (\lambda_i-\lambda_\eta )}\Bigg[\mho \Bigg\{\lambda_g e^{-\frac{\lambda_\eta +\lambda_i+\lambda_h \gamma_{th}}{\lambda_g \lambda_h}}\bigg[e^{\frac{\lambda_i}{\lambda_g \lambda_h}} \bigg\{\nonumber\\
 {}&\lambda_\eta  \lambda_i (\lambda_\eta +\lambda_h \gamma_{th}) \text{Ei}\left(\frac{\lambda_\eta }{\lambda_g \lambda_h}\right)-\lambda_\eta  \lambda_i \bigg(\lambda_\eta  +\lambda_h \gamma_{th}\bigg)\nonumber\\
 {}& \text{Ei}\left(\frac{\gamma_{th} \lambda_h+\lambda_\eta }{\lambda_g \lambda_h}\right)+\lambda_g\lambda_h^2 (\lambda_\eta -\lambda_i) (\lambda_g+\gamma_{th})e^{\frac{\lambda_\eta }{\lambda_g \lambda_h}}\bigg\}\nonumber\\
 {}&+\lambda_\eta  \lambda_i \left(-e^{\frac{\lambda_\eta }{\lambda_g \lambda_h}}\right) (\lambda_i+\lambda_h \gamma_{th}) \text{Ei}\left(\frac{\lambda_i}{\lambda_g \lambda_h}\right)\nonumber\\
 {}&+\lambda_\eta  \lambda_ie^{\frac{\lambda_\eta }{\lambda_g \lambda_h}} (\lambda_i+\lambda_h \gamma_{th}) \text{Ei}\left(\frac{\gamma_{th} \lambda_h+\lambda_i}{\lambda_g \lambda_h}\right)\bigg]\nonumber\\
 {}&+\lambda_g^3 \lambda_h^2 (\lambda_i-\lambda
   \eta )\Bigg\}\Bigg]\Bigg].\label{out_gamma_k_eta}
 \end{align}
\subsubsection*{Proof } \textit{See Appendix~C.}

\begin{proposition}
\textcolor{black}{As $\SNR\rightarrow \infty$, the rate outage probability for the perfect timing synchronization, imperfect SIC, and perfect CSI becomes}
\end{proposition}
\begin{align}
\textcolor{black}{
 \lim_{\SNR \to \infty}\tilde{P}_{out}^\eta\sim  \tilde{g}(\SNR)+ 2\lambda_{g}^2[g(\SNR)]^2,\label{AsPout_k3f}
}
\end{align}
\textcolor{black}{
where $\tilde{g}(\SNR)=\frac{\lambda_h g(\SNR)}{\Phi_k-(\Phi_\eta+\sum\limits_{i=1}^{k-1}\Phi_i)( 2^{\Upsilon}-1)}$ and $\Phi_\eta$ is the coefficient of power received due to ipSIC.}
\subsubsection*{Proof } \textit{See Appendix~D.}\\
\textcolor{black}{
Compared to the ideal case in \eqref{AsPout_k2f}, we observe that the adverse effect of $\Phi_\eta$ introduced by the imperfect SIC is to increase the outage probability.}


\begin{lemma}
The outage probability for the imperfect timing synchronization, perfect SIC, and perfect CSI is given as
\end{lemma}
\begin{align}
P_{out}^\varepsilon &={} \frac{1}{\Gamma (\alpha
   )}\Bigg[\left(-\frac{1}{\beta^2}\right)^{-\alpha } \beta ^{-\alpha} 
   \Bigg((-1)^{\alpha } \left(\frac{1}{\beta }\right)^{\alpha } \delta
   (\gamma_{th})\nonumber\\
   {}&+\left(-\frac{1}{\beta }\right)^{\alpha } \Bigg) \left(\Gamma
   (\alpha )-\Gamma \left(\alpha ,\frac{\gamma_{th}}{\beta }\right)\right)\Bigg],
   \label{out_gamma_k_varepsilon}
\end{align}
where $\alpha$ and $\beta$ are given in \eqref{pdf_iT}.
\subsubsection*{Proof } \textit{See Appendix~E.}

\begin{proposition}
\textcolor{black}{As $\SNR \rightarrow\infty$, the rate outage probability for the imperfect timing synchronization ($0<\varepsilon_1<1$), perfect SIC, and perfect CSI becomes}
\end{proposition}
\begin{align}
\textcolor{black}{
 \lim_{\SNR \to \infty}\tilde{P}_{out}^{\varepsilon} \sim  \frac{\lambda_h g(\SNR)}{\Phi_k-\sum\limits_{i=1}^{k-1}\Phi_i( 2^{\Upsilon}-1)}+ g^{\varepsilon}(\SNR).\label{AsPout_epif}}
\end{align}
\textcolor{black}{where $g^{\varepsilon}(\SNR)=2\varepsilon_1\lambda_{g}^2\left(\frac{2^{\Upsilon}-1}{\SNR}\right)^2$. 
 \subsubsection*{Proof } \textit{See Appendix~F.}\\
From this proposition, it is clear that the second term in \eqref{AsPout_k2f} quantifies the effect of the timing error on the outage probability.}

\begin{lemma}
The outage probability for the imperfect timing synchronization, imperfect SIC, and perfect CSI is given as
\end{lemma}
\begin{align}
P_{out}^{\varepsilon,\eta} &={} \frac{1}{\Gamma (\theta
   )}\Bigg[\left(-\frac{1}{\phi ^2}\right)^{-\theta } \phi ^{-\theta}\Bigg((-1)^{\theta } \left(\frac{1}{\phi }\right)^{\theta } \delta(\gamma_{th})\nonumber\\
   {}&+\left(-\frac{1}{\phi }\right)^{\theta } \Bigg) \bigg(\Gamma
   (\theta )-\Gamma \left(\theta ,\frac{\gamma_{th}}{\phi }\right)\bigg)\Bigg],
   \label{out_gamma_k_epi}
\end{align}
where the values of $\theta$ and $\phi$ are given in \eqref{pdf_gamma_eta_epi}.
\subsubsection*{Proof } \textit{See Appendix~G.}\\
\textcolor{black}{We note that Lemmas 1, 2, and 3 are special cases of this lemma. The following asymptotic analysis provides the corresponding limiting rate outage for $\SNR\rightarrow \infty$.}
 \begin{proposition}
\textcolor{black}{As $\SNR \rightarrow\infty$, the rate outage probability for the imperfect timing synchronization, imperfect SIC, and perfect CSI becomes}
\end{proposition}
\begin{align}
\textcolor{black}{
  \lim_{\SNR \to \infty}\tilde{P}_{out}^{\varepsilon,\eta} \sim  \tilde{g}(\SNR)+ g^{\varepsilon}(\SNR).
  }
  \label{AsPout_epi_etaf}
\end{align}
\subsubsection*{Proof } \textit{See Appendix~H.}\\
\textcolor{black}{In fact, this result in \eqref{AsPout_epi_etaf} is in line with the composite degradations found in Propositions 2 and 3. In the following lemma and proposition, we will investigate the impact of the imperfect CSI.
}

\begin{lemma}
\textcolor{black}{The outage probability for the perfect timing synchronization, perfect SIC, and imperfect CSI is given as}
\end{lemma}
\textcolor{black}{
\begin{align}
P_{out}^{\chi}= {} & \frac{1}{\lambda_\chi^2 \lambda_h^2 }\sum\limits_{i=1}^{I}\Bigg[\textcolor{black}{\mho} e^{-\frac{\lambda_i+\lambda_h \gamma_{th}}{\lambda_\chi \lambda_h}} \Bigg(\lambda_i (\lambda_i+\lambda_h \gamma_{th}) \text{Ei}\left(\frac{\lambda_i}{\lambda_\chi \lambda_h}\right)\nonumber\\
&-\lambda_i (\lambda_i+\lambda_h \gamma_{th}) \text{Ei}\left(\frac{\gamma_{th} \lambda_h+\lambda_i}{\lambda_\chi \lambda_h}\right)+\lambda_\chi \lambda_h e^{\frac{\lambda_i}{\lambda_\chi \lambda_h}}\nonumber\\
&
   \bigg(\left(e^{\gamma_{th}/\lambda_\chi}-1\right) (\lambda_\chi \lambda_h+\lambda_i)-\lambda_h \gamma_{th}\bigg)\Bigg)\Bigg].
   \label{out_gamma_k_chi}
\end{align}
} 
\subsubsection*{Proof } \textit{See Appendix~I.}\\

 \begin{proposition}
\textcolor{black}{As $\SNR \rightarrow\infty$, the outage probability for the perfect timing synchronization, perfect SIC, and imperfect CSI becomes}
\end{proposition}
\textcolor{black}{
\begin{align}
 \lim_{\SNR \to \infty}\tilde{P}_{out}^\chi   \sim   \frac{\lambda_h g(\SNR)}{\Phi_k-\sum\limits_{i=1}^{k-1}\Phi_i( 2^{\Upsilon}-1)}+ 2\lambda_{\chi}^2[g(\SNR)]^2,
 \label{AsPout_k6f}
\end{align}
where $\lambda_\chi =1/ \sigma_\chi^2 = 1/(\sigma_{\varrho}^2+\rho^2\sigma_{g}^2)$.
}
\subsubsection*{Proof } \textit{See Appendix~J.}\\
\textcolor{black}{
Since the outage rates are obtained in closed-form expressions in the five lemmas, it is possible to estimate how the outage performance of the STBC-CNOMA scheme changes as various system parameters change. In addition, the asymptotic analysis in the propositions provide the insights into how the rate outage is degraded by the three imperfections. Our analysis in the lemmas and propositions will be validated by comparing with simulation results in Section VI.}

\section{Complexity Analysis}
\label{sec:complexity_analysis}
In this section, we will compare the complexity of STBC-aided cooperative NOMA (STBC-CNOMA) with other cooperative NOMA techniques in terms of the number of SIC performed, which is widely used to evaluate the computational complexity of NOMA as in~\cite{complx1,complx2,complx3,complx4,complx5}. We consider five different NOMA schemes:  conventional cooperative NOMA~(CCN)~\cite{ZDCNOMA}, cooperative relay systems using NOMA~(CRS-NOMA)~\cite{CRS-NOMA}, CRS-NOMA novel design~(CRS-NOMA-ND)~\cite{CRS-NOMA-ND}, cooperative relay selection by using STBC (CRS-STBC-NOMA)~\cite{kader2016cooperative}, and lastly STBC-CNOMA. We note that this is the first extensive comparison of the five schemes, which quantifies the total number of SIC performed for a given number of users $K$.

The total number of SIC performed by CCN is given by  
\begin{equation}\label{SIC_cNOMA}
     SIC_{ccn} =\sum\limits_{j=0}^{K-2}\Bigg[ \sum\limits_{i=1}^{K-1-j}{(K-(i+j))}\Bigg].
\end{equation} 
\textcolor{black}{Also, the total number of SIC performed by CRS-NOMA, which uses a half-duplex relay between the source and each user, is given by} 
\begin{equation}\label{CRS_NOMA}
     SIC_{crs-noma} =\sum\limits_{j=1}^{K}j.
\end{equation}
\textcolor{black}{Also, the total number of SIC performed by CRS-STBC-NOMA, which has a two-phase communication from the source to each user by means of $2 \times 1$~STBC, is given by}  
\begin{equation}\label{CRS_STBC_NOMA}
     SIC_{crs-stbc-noma} =\sum\limits_{j=1}^{K}4j.
\end{equation} 
In this scheme, the source is equipped with two transmit antennas, and each relay is equipped with one receive antenna and two transmit antennas, whereas each user is equipped with one receive antenna. 

On the other hand, the  total number of SIC performed by CRS-NOMA-ND is given by  
\begin{equation}\label{CRS_NOMA_ND}
     SIC_{crs-noma-nd} =\sum\limits_{j=1}^{K}2j.
\end{equation}
In this scheme, two sources transmit two symbols to two users by means of superposition coding, and each user decodes its symbol by MRC and SIC.  Whereas, the total number of SIC performed by STBC-CNOMA is given by 
\begin{equation}\label{SIC_STBCNOMA}
     SIC_{stbc-cnoma} =\sum\limits_{i=1}^{K-1} {(K-i)}.
\end{equation}
It is noted that \eqref{SIC_cNOMA} and \eqref{SIC_STBCNOMA} show that for a larger number of users, complexity of CCN increases due to higher number of SIC. Whereas, the number of SIC is not increased in cooperation phase of STBC-CNOMA. Therefore, it achieves the diversity gain of STBC codes and maintains the number of SIC same as that of conventional direct NOMA scheme.

\begin{table}[t!]
\caption{\textcolor{black}{A comparison of time slots required and number of transmissions for different cooperative NOMA schemes.}}
\label{Table4}
\centering
\begin{tabular}{|P{3cm}|P{1.3cm}|P{1.8cm}|}
\hline
\cline{1-2}
Algorithm & No. of Time Slots& No. of Transmissions \\
\hline\hline
CCN~\cite{ZDCNOMA}   & $K$ & $K$  \\
\hline

CRS-NOMA~\cite{CRS-NOMA}   & $K$& $K$ \\
\hline
CRS-NOMA-ND~\cite{CRS-NOMA-ND}  & $K$& $K$ \\
\hline
CRS-STBC-NOMA~\cite{kader2016cooperative}  & $2K$& $4K$ \\
\hline
STBC-CNOMA   & $K-1$& $2K-3$  \\
\hline
\end{tabular}
\end{table}


In Table \ref{Table4}, we compare the number of time slots required for the complete transmission in various cooperative schemes for even $K\geq4$. As shown in the table, STBC-CNOMA requires $K-1$ time slots for the complete transmission. Therefore, saving one time slot as compared to that of in CCN  and CRS-NOMA. This time slot can be utilized in other types of signaling, which makes STBC-CNOMA more efficient scheme. \textcolor{black}{In addition, in the last column of Table \ref{Table4}, we compare the number of transmissions, which indicates the communication overhead. As shown in the table, CRS-STBC-NOMA has the largest number of transmissions. Due to the STBC phase, we can find out that STBC-CNOMA also requires more number of transmissions compared to the other three scheme for any $K\geq4$.}

\begin{figure}[t]
  \begin{center}
  \includegraphics[scale=0.5]{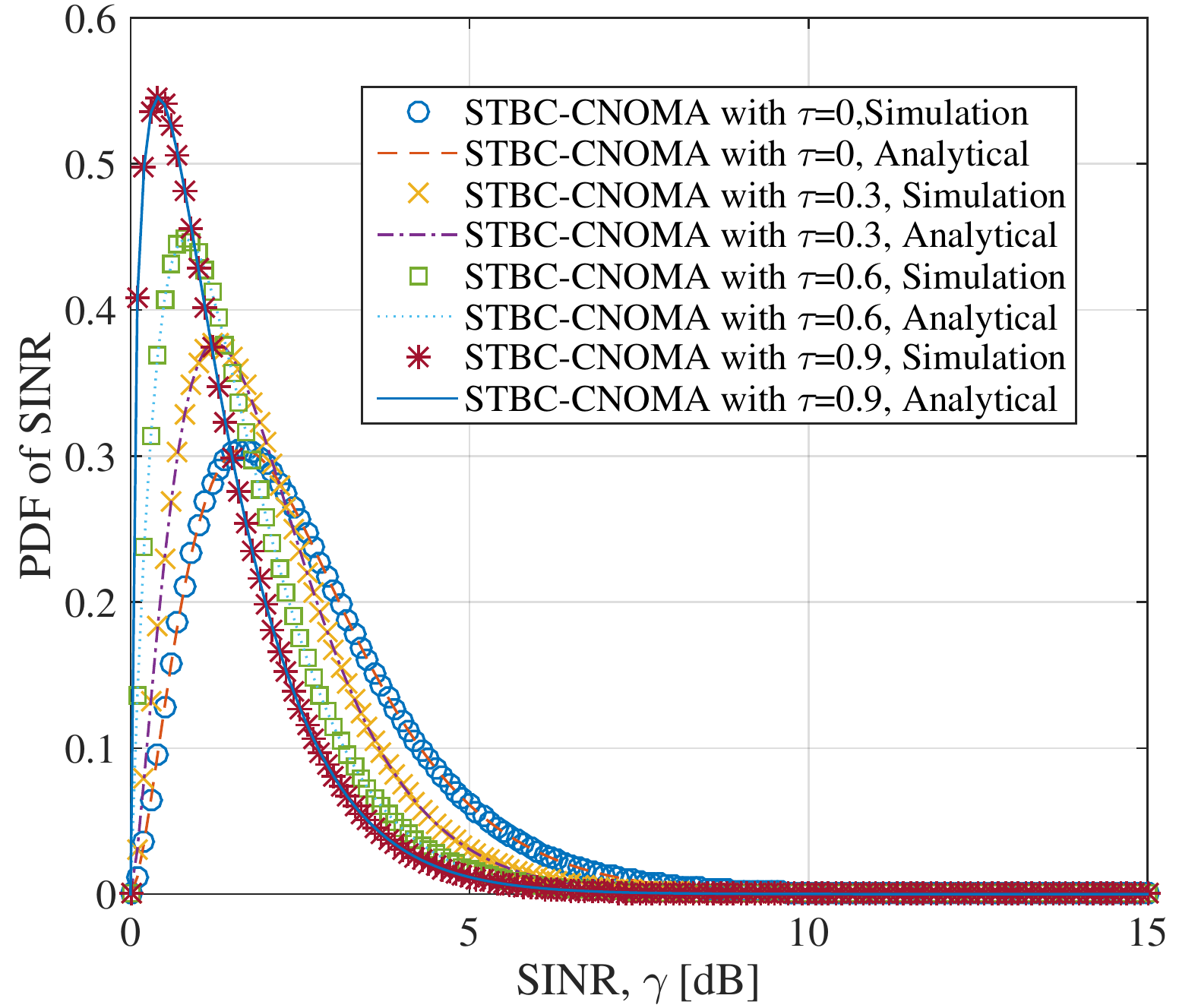}
  \caption{The SINR PDFs for the perfect SIC, perfect CSI, and variable timing offsets, when $K=4$.}\label{n1}
  \end{center}
\end{figure}

\section{Simulation Results}
\label{sec:simulation_results}
In this section, we present numerical and simulation results and validate our analysis in the previous sections. 
We consider a downlink NOMA system with one BS and $K$ users. 
As in~\cite{ ZDCNOMA, CRS-NOMA, jamal2018efficient}, we assume that the channels between the BS and each user, and inter-users are flat fading Rayleigh channels. The noise power spectral density is considered as $-174$~dBm/Hz. The rate threshold $\Upsilon$ is set to be 2 bits per channel use~(BPCU).
Each user is considered as stationary. The  parameters $\zeta_h$, $\zeta_\eta$, $\zeta_{g_{k,k-2}}$, $\zeta_{g_{k,k-3}}$ and \textcolor{black}{$\zeta_\chi$} are considered to be unity. Symbol duration $T$ is also considered as unity for the simplicity. \textcolor{black}{For the STBC-CNOMA scheme, the transmit power at the BS~(i.e., $P_{NOMA}$) is considered as 45 dBm, whereas the power transmitted during STBC cooperation phase~(i.e., $p_s$) is considered as half of the power transmitted from the BS. Power coefficients for simulations with $K=4$ are $\Phi_1=0.1$, $\Phi_2=0.2$, $\Phi_3=0.3$, and $\Phi_4=0.4$ for $k\in\{1, 2, 3, 4\}$, respectively. It is to be noted that we assume the same total power budget for all of the schemes in the simulation for the fair comparison.}

Fig.~\ref{n1} provides the comparison of simulation and analytical results for the SINR PDFs for different timing offsets, assuming $K=4$ with the perfect SIC and \textcolor{black}{CSI}. \textcolor{black}{It can be noticed that the simulation results closely match the analytical results for  
various timing offsets $\tau\in\{0, 0.3, 0.6, 0.9\}$. Thus, as an essential component for further performance analysis, the SINR PDFs derived in Section VI have been validated. Also, as expected, as the timing offset increases, the mean of the distribution approaches zero. In other words, the average SINR of the user decreases, as the timing offset~$\tau$ increases, which means that the outage probability of the user is an increasing function of $\tau$.}

\begin{figure}[t]
  \begin{center}
   \includegraphics[scale=0.5]{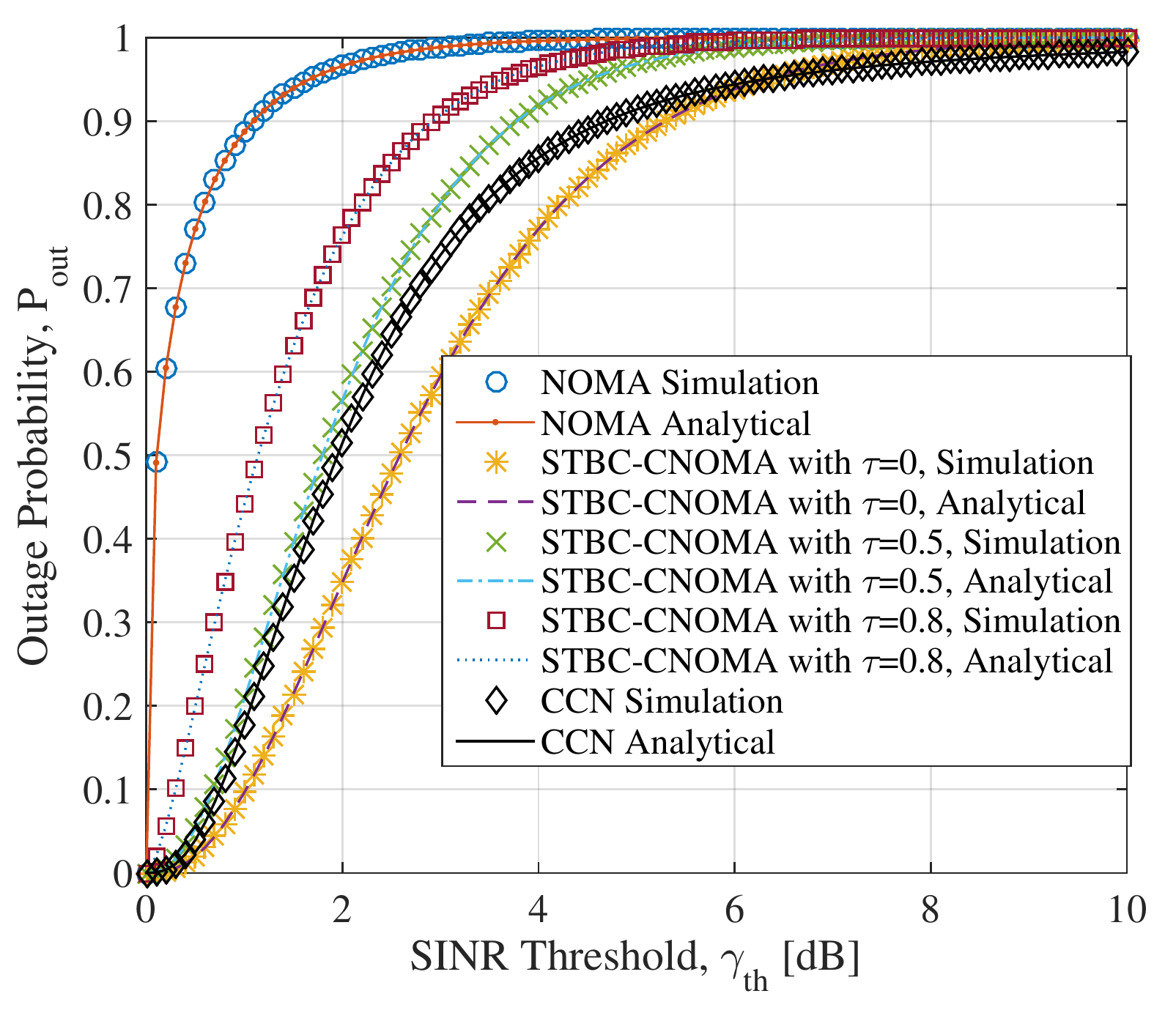}
  \caption{Outage probability performance for the perfect SIC, perfect CSI, and variable timing offsets, when $K=4$.}\label{n4}
  \end{center}
\end{figure}

In Fig.~\ref{n4}, we compare the outage probabilities of NOMA, CCN, and STBC-CNOMA for $\tau=\{0, 0.5, 0.8\}$ and $K=4$. In the figure, the horizontal axis represents the SINR threshold $\gamma_{th}$. \textcolor{black}{The simulation and analytical results show great correlation with each other, which validates Lemmas 1 and 3}. Further, in the figure, we observe that the outage probability of NOMA is the highest for a given $\gamma_{th}$, because it is a non-cooperative scheme that does not provide diversity gain. It is also noted that STBC-CNOMA outperforms the CCN for the low SINR thresholds. Also, the performance of the STBC-CNOMA with $\tau$ = 0.5 is similar to that of CCN for SINR threshold of up to 2 dB. Fig.~\ref{n4} also demonstrates that the outage probability of the STBC-CNOMA approaches to that of NOMA with the timing offset, $\tau$ approaching to 1. Therefore, for the low SINR threshold and $\tau< 0.5$, STBC-CNOMA is still an attractive scheme.

\begin{figure}[t]
  \begin{center}
    \includegraphics[scale=0.5]{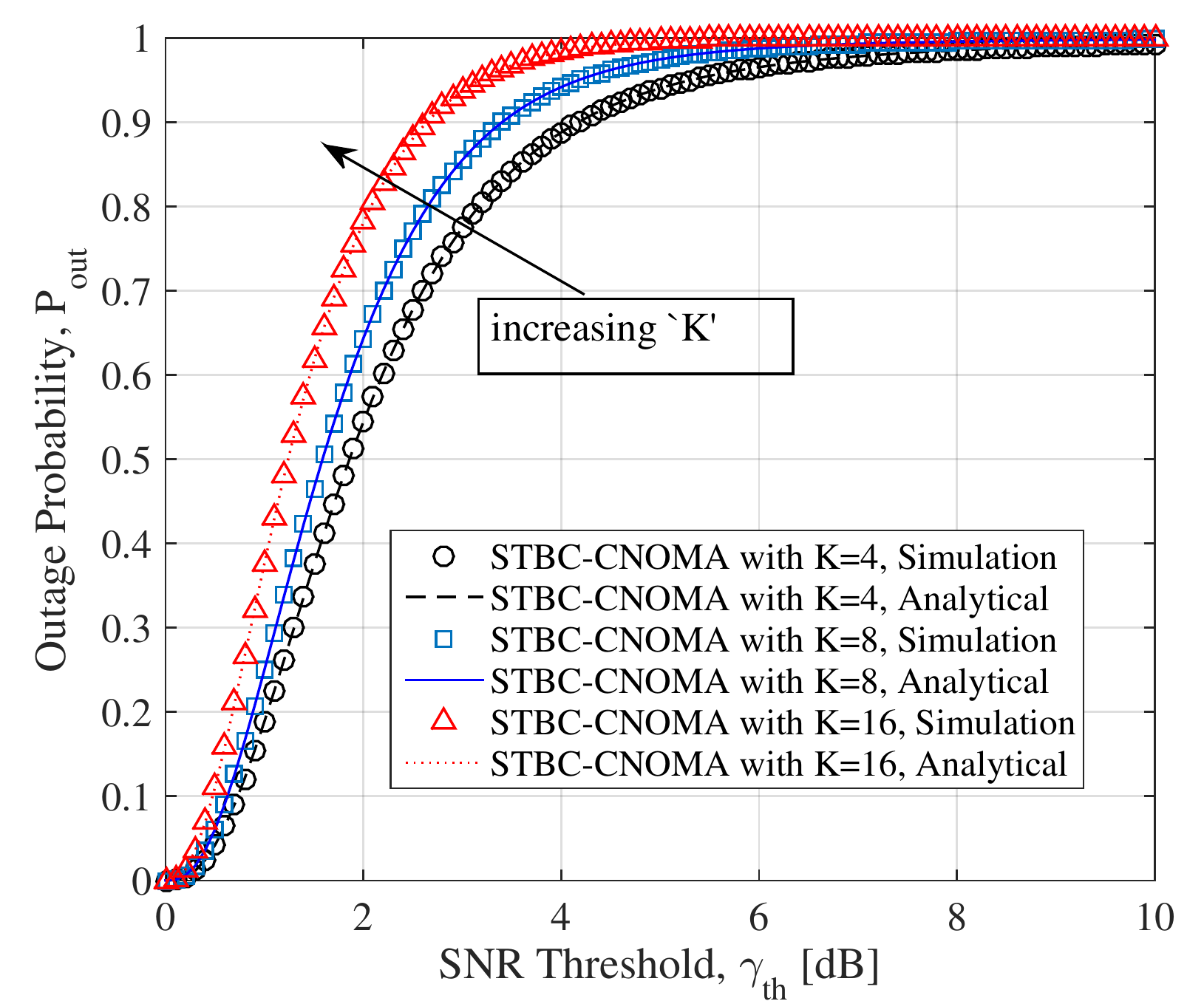}
  \caption{Outage probability performance for the perfect timing,  perfect CSI, and imperfect SIC = $-5$dBs, when $K\in\{4, 8, 16\}$.}\label{n3}
  \end{center}
\end{figure}

\begin{figure}[t]
  \begin{center}
    \includegraphics[scale=0.5]{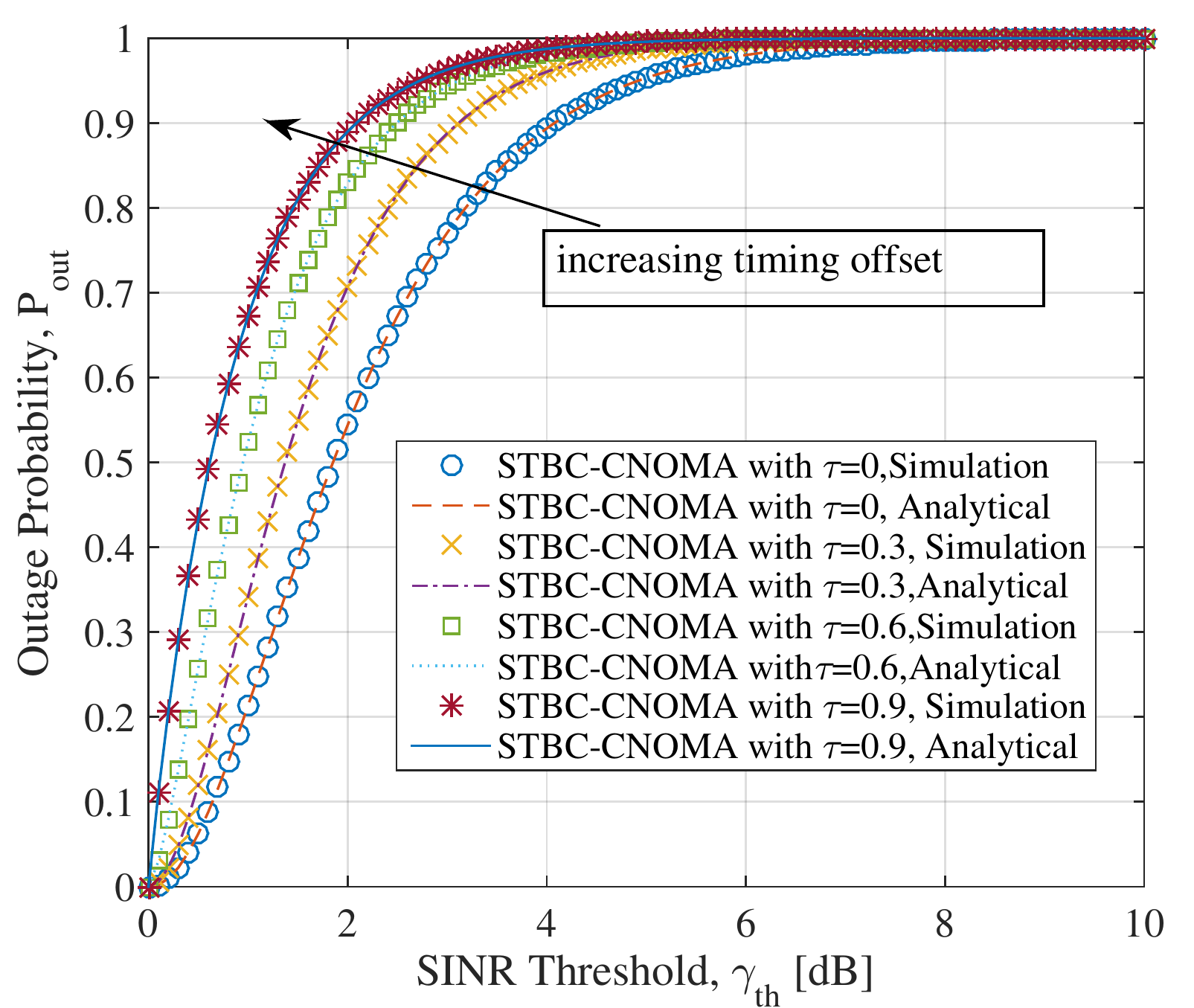}
  \caption{Outage probability performance with $K=4$ for the imperfect SIC~=$-5$~dBs, perfect CSI, and variable timing offsets.}\label{n5}
  \end{center}
\end{figure}

\begin{figure}[t]
  \begin{center}
    \includegraphics[scale=0.65]{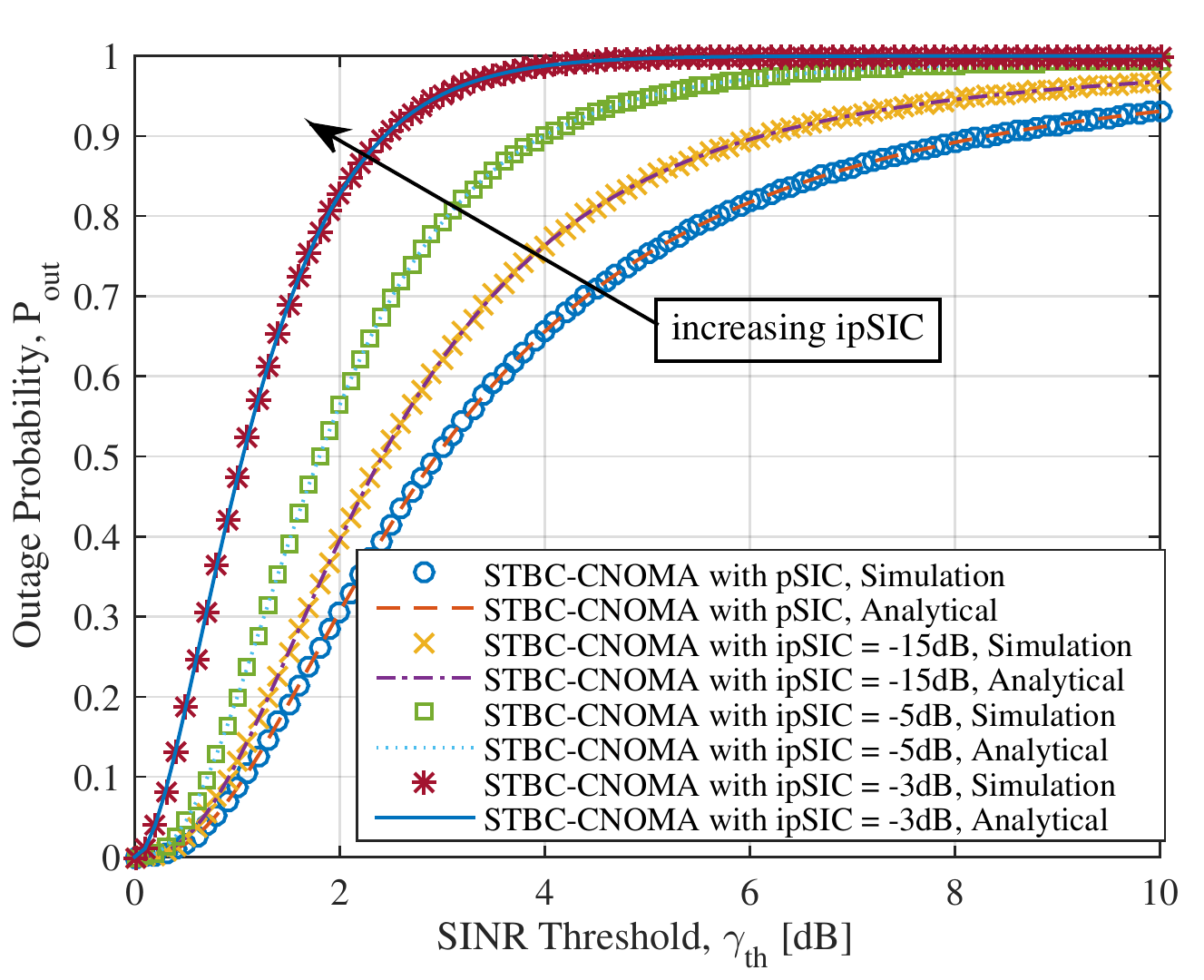}
  \caption{\textcolor{black}{Outage probability of User 4 in STBC-CNOMA with the imperfect SIC, perfect timing synchronization, and perfect CSI.}}\label{OUT4_ipSIC_pTiming_06}
  \end{center}
\end{figure}

Fig.~\ref{n3} shows both analytical and simulation results of the outage performance of the STBC-CNOMA with the total number of users $K\in\{4, 8, 16\}$. In the figure, we observe the outage performance degradation with the increasing number of users. Further, it is noted that there is 
not much difference in the outage performance as $K$ exceeds 8. In other words, in the absence of any imperfection, the outage performance of STBC-CNOMA is not much degraded with the increasing number of users. Also, the analytical and simulation results closely match, which validates Lemma 2 in Section IV. 

\begin{figure}[t]
  \begin{center}
    \includegraphics[scale=0.65]{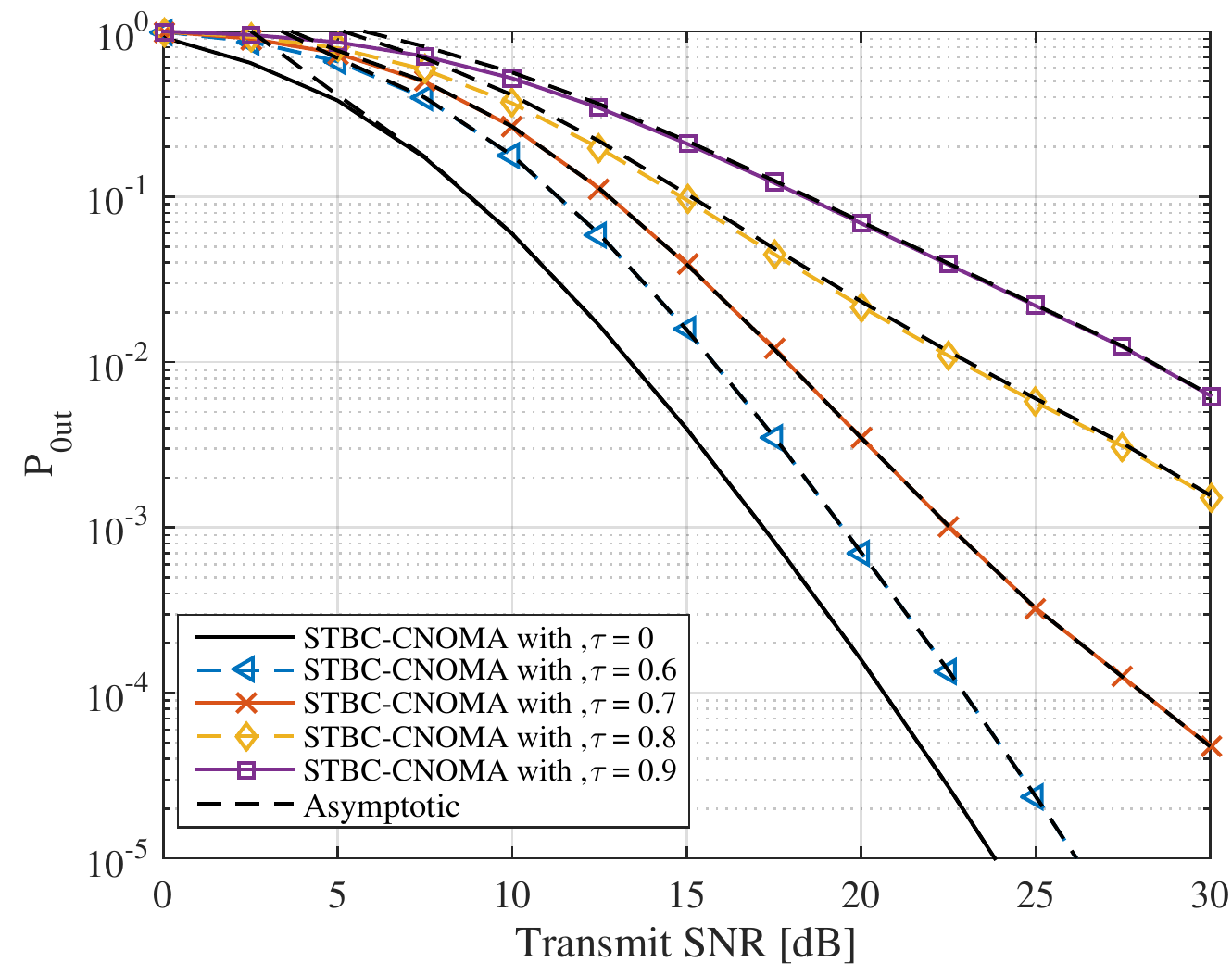}
  \caption{\textcolor{black}{Rate outage probability of User 4 in STBC-CNOMA at $\tau = \{0, 0.6, 0.7, 0.8, 0.9\}$, perfect CSI, and perfect SIC.}}\label{OUT4_pSIC_ipTiming}
  \end{center}
\end{figure}

\begin{figure}[t]
  \begin{center}
    \includegraphics[scale=0.75]{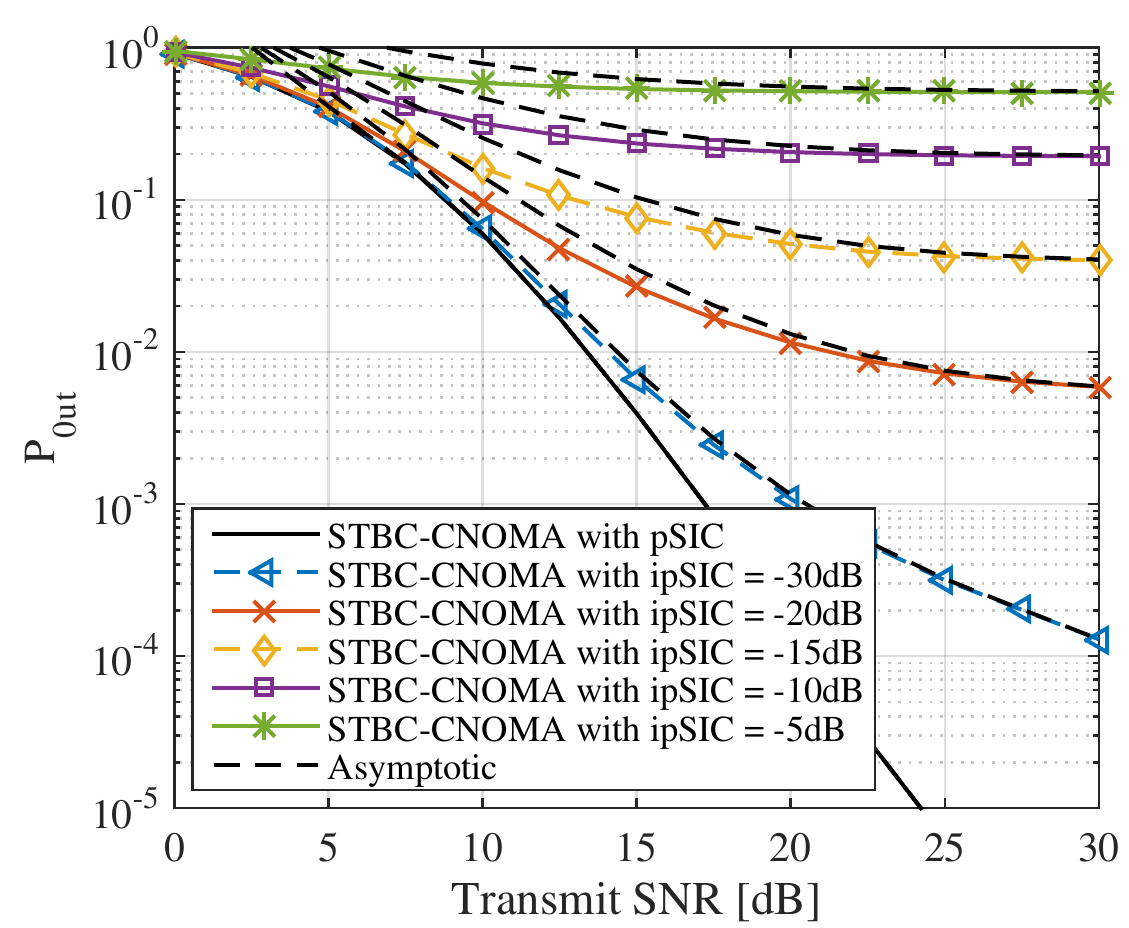}\caption{\textcolor{black}{Rate outage probability of User 4 in STBC-CNOMA with the perfect timing synchronization, perfect CSI, and imperfect SIC.}}\label{OUT4aaaa_ipSIC_pTiming}
  \end{center}
  \end{figure}
  
\begin{figure}[t]
  \begin{center}
    \includegraphics[scale=0.75]{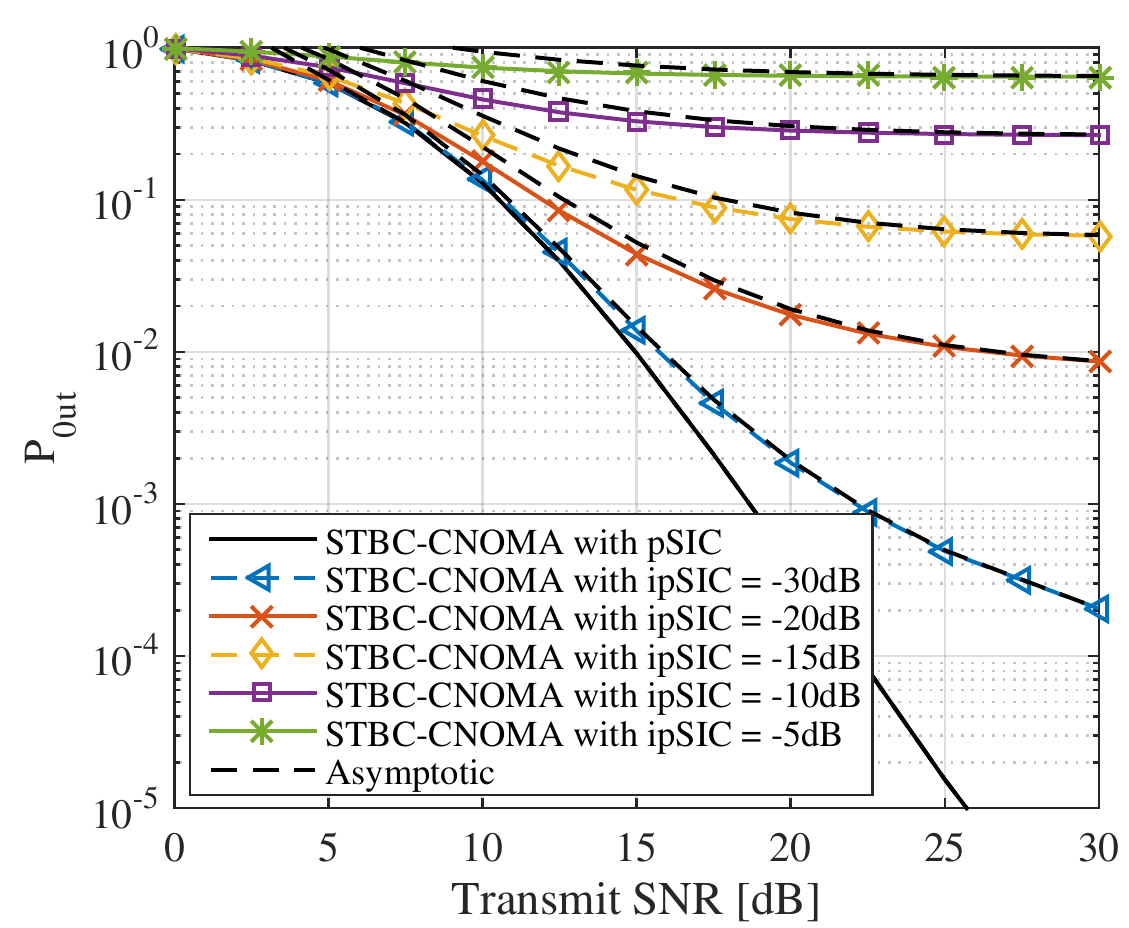}
  \caption{\textcolor{black}{Rate outage probability of User 4 in STBC-CNOMA with the timing offset of 0.5, perfect CSI, and imperfect SIC.}}\label{OUT4_ipSIC_ipTiming_05}
  \end{center}
\end{figure}

\begin{figure}[t]
  \begin{center}
    \includegraphics[scale=0.75]{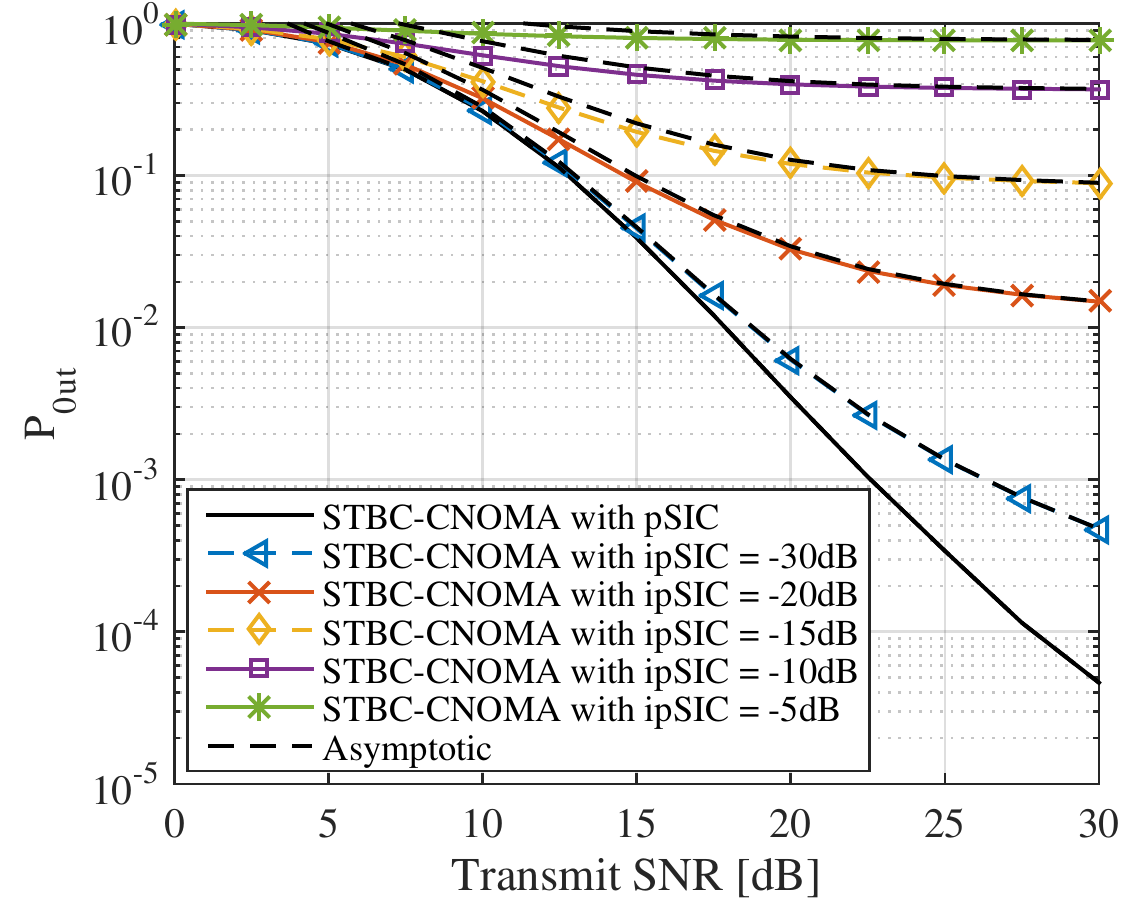}
  \caption{\textcolor{black}{Rate outage probability of User 4 in STBC-CNOMA with the timing offset of 0.7, perfect CSI, and imperfect SIC.}}\label{OUT4_ipSIC_ipTiming_07}
  \end{center}
\end{figure}

Similarly, Fig.~\ref{n5} shows the analytical and simulation results for the performance of STBC-CNOMA with different timing offsets for the perfect CSI and imperfect SIC of $-5$~dBs, when $K=4$. \textcolor{black}{Fig~\ref{OUT4_ipSIC_pTiming_06} depicts the outage probability with different levels of the SIC imperfection. In this case, Fig.~\ref{n5} and Fig~\ref{OUT4_ipSIC_pTiming_06} show the great correlation between the simulation and analytical results based on Lemma 4 in Section IV. Also, as shown in the figures, we observe that the outage probability increases sharply, as $\tau$ and the SIC imperfection increase. 
} 
 
Fig.~\ref{OUT4_pSIC_ipTiming} presents a comparative analysis of the rate outage performance as function of the transmit SNR, $\SNR$, of User 4 for STBC-CNOMA with the perfect SIC and different values of timing offsets. We assume that the rate threshold is 2 bits per channel use (BPCU) and there are 4 users (i.e., $K=4$) in the system. \textcolor{black}{Also, the dotted lines correspond to the asymptotic rate outage probabilities derived in Section IV.} The results show that the rate outage degrades with the increase in the timing offset $\tau$, which is in line with  the previous simulation results. The user has the best outage performance for $\varepsilon_1=1$, i.e., when there is no timing offset. However, in case that the users are not perfectly synchronized, the orthogonality of the received symbols is compromised, which leads to performance degradation. The user with $\uptau=1$ experiences an outage probability close to that of non-cooperative NOMA. \textcolor{black}{We also observe the asymptotic analysis curves based on Propositions 1 and 3, which show good agreement with both simulation and original analytical results in the high SNR regime.}

\begin{figure}[t]
  \begin{center}
    \includegraphics[scale=0.65]{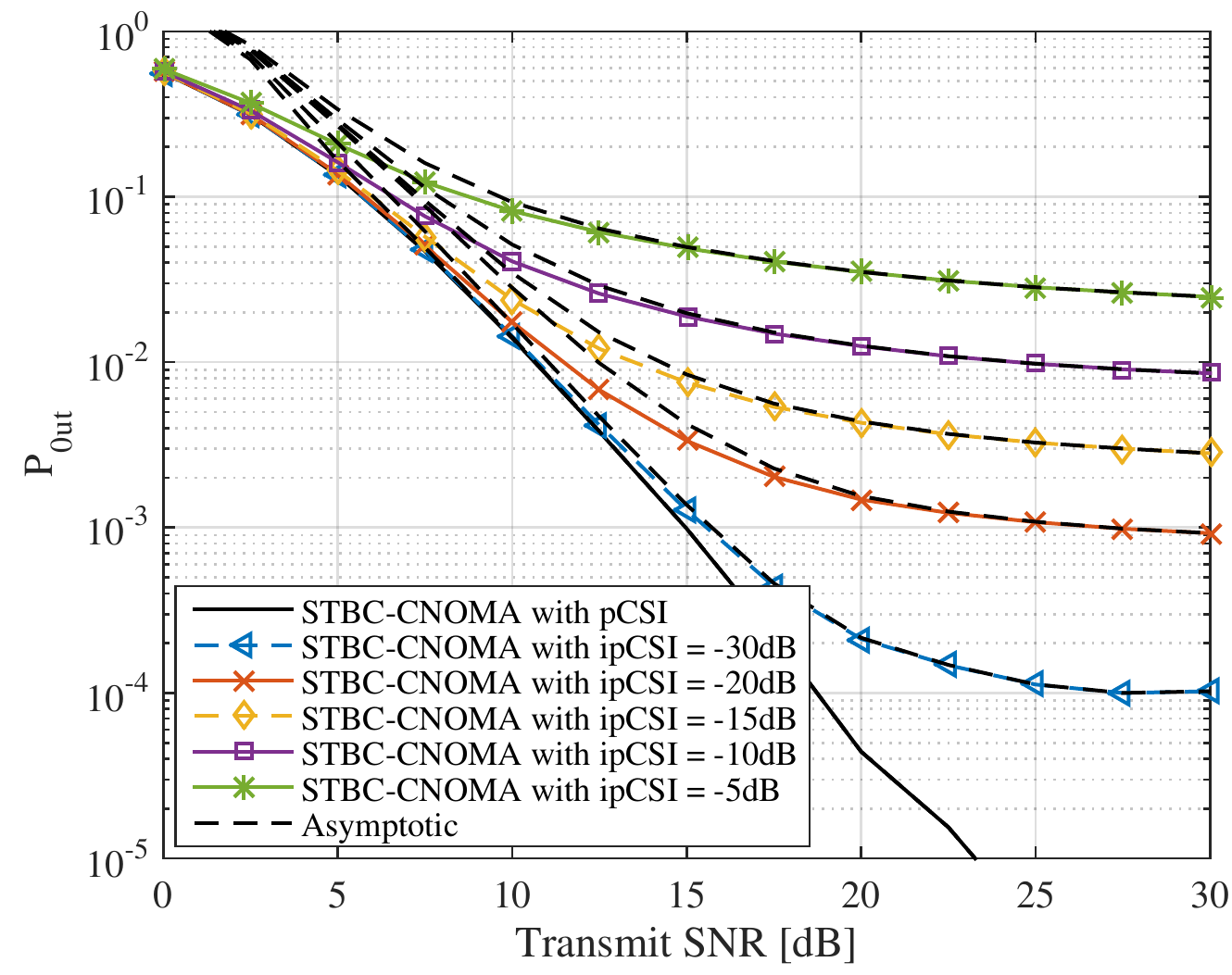}
  \caption{\textcolor{black}{Rate outage probability of User 4 in STBC-CNOMA with the perfect timing synchronization, perfect SIC, and imperfect CSI.}}\label{OUT4_pSIC_pTiming_ipCSI}
  \end{center}
\end{figure}
\begin{figure}[t]
  \begin{center}
    \includegraphics[scale=0.65]{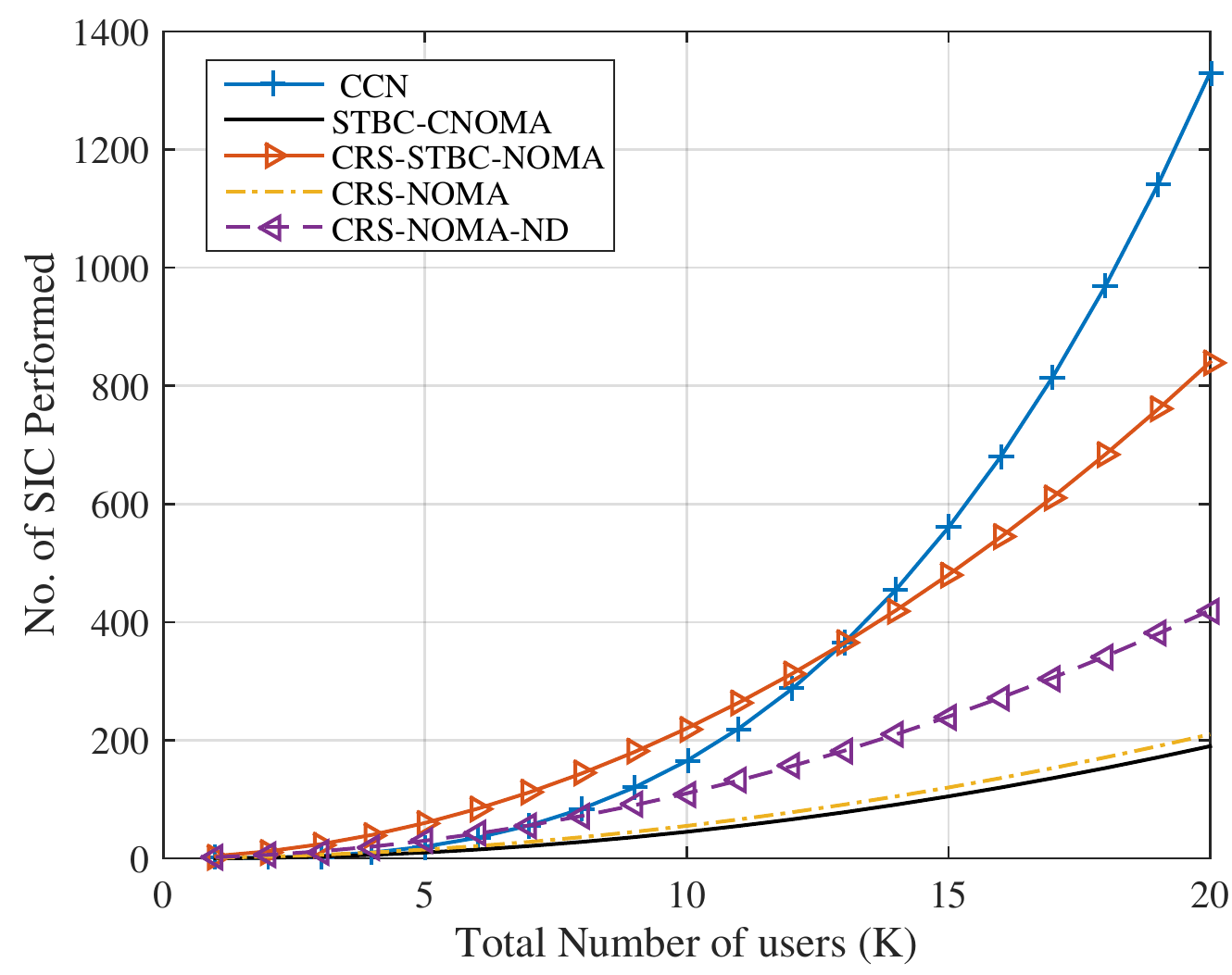}
  \caption{A  comparison of number of SIC performed in different flavors of cooperative NOMA.}\label{SIC_comparison}
  \end{center}
\end{figure}

Fig.~\ref{OUT4aaaa_ipSIC_pTiming} shows the rate outage probability of User 4 with the perfect timing synchronization, perfect CSI, and imperfect SIC. In the figure, we can find that the performance of User 4 degrades significantly with the increasing impact of imperfect SIC even with perfect timing synchronization. This implies that even if the system is perfectly synchronized but with imperfect SIC, its performance is not reliable. Fig.~\ref{OUT4_ipSIC_ipTiming_05} gives a snapshot of rate outage probability of User 4 with the imperfect SIC and timing offset $\tau$ of $0.5$. As expected, the figure shows that the outage probability of User 4 degrades severely with the increasing effect of imperfect SIC. It is obvious that a small increase in the intensity of imperfect SIC leads to a higher degradation in the performance of the system. \textcolor{black}{Furthermore, in both Figs. \ref{OUT4aaaa_ipSIC_pTiming} and \ref{OUT4_ipSIC_ipTiming_05}, as $\SNR$ increases, we observe the asymptotic analysis curves show the almost identical results as the simulation and exact analysis, which validates Propositions 2 and 4. Fig.~\ref{OUT4_ipSIC_ipTiming_07} illustrates the rate outage probability with the timing offset $\tau$ of 0.7 and the imperfect SIC. As expected, the rate outage probability in Fig.~\ref{OUT4_ipSIC_ipTiming_07} is higher compared to Fig.~\ref{OUT4_ipSIC_ipTiming_05}, which corresponds to $\tau=0.5$, for the same SIC condition.}

 \textcolor{black}{Fig.~\ref{OUT4_pSIC_pTiming_ipCSI} shows the effect of the channel estimation error (i.e., CSI imperfection) on the rate outage performance of STBC-CNOMA. In the figure, we first observe that the analytical results indicated by the solid lines have excellent correlation with the corresponding simulation results, which are indicated by the different markers. In addition, the dotted lines, which correspond to the asymptotic analysis, approach to the solid lines and markers, as $\SNR$ increases. Thus, both Lemma 5 and Proposition 5 have been validated. Also, in the figure, the performance degrades rapidly with the increase in the magnitude of the imperfection in the channel estimation. By comparing the results in Figs. \ref{OUT4aaaa_ipSIC_pTiming} and \ref{OUT4_pSIC_pTiming_ipCSI}, it is observed  that the impact of the imperfect SIC is greater compared to that of the imperfect CSI. This is due to the fact that the SIC imperfection degrades the SINR performance in direct NOMA phase as well as in cooperative NOMA phase, because accurate detection of weak users’ symbols in direct NOMA phase depends on the level of perfection in SIC. The imperfection in the detected weak users’ symbols leads to decrease in SINR in cooperative NOMA phase.} In addition, comparing the results in Figs.~\ref{OUT4_pSIC_ipTiming} and \ref{OUT4_pSIC_pTiming_ipCSI}, we can conclude that the system with the perfect SIC and the timing offset of 0.7 performs better than the system with the imperfect SIC of $-30$ dB and perfect timing synchronization. This shows that the impact of the imperfect SIC is more significant compared to that of imperfect timing synchronization or imperfect CSI. 

 Lastly, Fig.~\ref{SIC_comparison} shows the complexity comparison in Section V. As shown in the figure, it is obvious that the CCN becomes computationally expensive for higher number of users due to its exponential increase in the number of SIC to be performed. On the other hand, STBC-CNOMA has substantially reduced the number of SIC. For example, when the total number of users is 6, the CCN requires 35 SIC whereas in STBC-CNOMA, the number of SIC required is 15, making a 57$\%$ reduction in number of SIC performed.  
This reduction of SIC increases to 72.72 $\%$ and 83.17$\%$ as the number of users increases to 10 and 18, respectively. We can say that the CRS-STBC-NOMA has the worst performance in terms of complexity for low and medium numbers of users. But, for higher number of users, i.e., $K>13$, conventional cooperative NOMA has the worst performance in terms of complexity. The proposed scheme, STBC-CNOMA, outperforms all of the other schemes in terms of numbers of SIC performed, which is desirable in hardware-limited networks.

\section{Conclusions}
\label{sec:conclusions}
In this paper, we have provided the theoretical framework to incorporate \textcolor{black}{three} realistic impairments, which are timing error, SIC imperfection, and \textcolor{black}{CSI impairment}, with outage performance. We have derived the closed-form expressions of the outage probabilities for the different combinations of the three impairments. Further, 
the complexity of STBC-CNOMA has been compared with existing cooperative NOMA protocols such as CCN, CRS-NOMA, CRS-STBC-NOMA and CRS-NOMA-ND in terms of the total number of SIC. Through both analysis and simulation, we show that STBC-CNOMA can be an attractive solution for systems with higher number of users or devices and having low power constraints. The simulation results also have shown that for a small number of users~(i.e., $K\leq4$) STBC-CNOMA without any imperfection outperforms CCN, until the SINR threshold exceeds a certain value. Even with moderate timing offset $\tau<0.5$, we have observed that the outage performance degradation of STBC-CNOMA relative to CCN is not significant. On the other hand, the impact of the imperfect SIC on the outage performance of STBC-CNOMA is more significant compared to those of the timing offset and \textcolor{black}{the imperfect CSI}. Therefore, considering the smaller number of SIC in STBC-CNOMA compared to the other cooperative NOMA protocols, we can conclude that STBC-CNOMA is an effective solution to achieve high reliability for the same SIC imperfection condition. 
\begin{appendices}
\renewcommand{\thesectiondis}[2]{\Alph{section}:}

\textcolor{black}{
\section{Proof of Lemma 1}
Suppose that $A\sim Exp(\lambda_h)$ and $B$ is the sum of exponential RVs resulting in a hypo-exponential RV, i.e., $B\sim hypoexp(\boldsymbol{\lambda_i})$ with $\boldsymbol{\lambda_i}$ is a vector given as $\boldsymbol{\lambda_i}=\{\lambda_1,\lambda_2,\lambda_3,\dots,\lambda_I\}$, as described in Section IV-A, then the PDF of $A$ is given by
\begin{align}
f_A(a) = \lambda_h e^{-\lambda_ha}, ~a\geq0.
 \label{pdf_a}
\end{align}
In case of only one interfering user, the PDF of B is given as 
\begin{align}
f_B(b) =  \lambda_1 e^{-\lambda_1b}, ~b\geq0.
 \label{pdf_b_i_0}
\end{align}
For more than one interferers, the PDF of B is given as
\begin{align}
f_B(b) =  \Bigg(\prod_{i=1}^{I} \lambda_i\Bigg)\Bigg[ \sum\limits_{i=1}^{I}\frac{e^{-\lambda_ib}}{ \prod_{j=1,i\neq j}^{I}(\lambda_j-\lambda_i)}\Bigg],
 \label{pdf_b_i_m}
\end{align}
where $b>0$ and $I$ is the total number of interferers. 
The PDF of  $Q_1=\frac{A}{B}$ can be expressed as
\begin{align}
f_{Q_1}(q) = \int_{0}^{\infty}bf_A(bq)f_B(b)db .
 \label{pdf_q1q}
\end{align}
By substituting ~\eqref{pdf_a} and ~\eqref{pdf_b_i_m} into ~\eqref{pdf_q1q}, we get the form in~\eqref{pdf_q}.
\textcolor{black}{Then, as in~\cite{prob1}, the PDF  of $Q_1 = \frac{A }{B}$ is given by} 
\begin{align}
f_{Q_1}(q) = {}&\frac{ \lambda_h\psi_1\left[\psi_2 + q\lambda_h \left( \sum\limits_{i=1}^{I}(i+1)q^{i-1}\lambda_h^{i-1} \psi^{I-i-1}\right) \right]} {\prod_{j=1}^{I}  (\lambda_hq+\lambda_j)^2}
,
 \label{pdf_q}
\end{align}
where $q>0$. Hence, using this PDF, the mean and variance of $Q_1$ can be obtained as
$\mathop{\E}[Q_1] = \frac{ \psi_{1} \psi_{3}} {\psi_{4}}$
and
$\mathop{\Var}[Q_1] = \frac{\lambda_h\psi_1 (\lambda_h\psi_1\psi_{6}^2+2 \psi_{7}\psi_{8} )} {\psi_{9}}$,
respectively. Similarly, for the perfect timing synchronization, let $Z=C+D$, which is the sum of two exponential RVs. Thus, $Z \sim Gamma(\lambda_{g})$, which corresponds to the following PDF of $Z$
\begin{align}
f_Z(z) = \frac{z}{\lambda_g^2} e^{-z\lambda_g},~z>0.
 \label{pdf_z}
\end{align}
\textcolor{black}{Without any imperfection, we can write \eqref{sinr_k_STBC-CNOMA} as $L=Q_1+Z$, where the PDF of $Q_1$ is given as \eqref{pdf_q}. By convolving the PDFs of $Q_1$ and $Z$, we obtain the PDF of $L=Q_1+Z$, which can be found in~\cite{prob1} as }
\begin{align}
f_L(l) = {} &\frac{1}{\lambda_g^3 \lambda_h^2}\sum\limits_{i=1}^{I}\Bigg[\mho e^{-\frac{\lambda_i+\lambda_h l}{\lambda_g \lambda_h}} \Bigg(\lambda_i (-\lambda_g
   \lambda_h+\lambda_i+\lambda_h l)\nonumber\\
{} & \left(\text{Ei}\left(\frac{l \lambda_h+\lambda_i}{\lambda_g
   \lambda_h}\right)-\text{Ei}\left(\frac{\lambda_i}{\lambda_g \lambda_h}\right)\right)+\nonumber\\
   &\lambda_g\lambda_h e^{\frac{\lambda_i}{\lambda_g \lambda_h}} \left(\lambda_i-\lambda_i
   e^{l/\lambda_g}+\lambda_h l\right)\Bigg)\Bigg],
 \label{pdf_pTpSIC}
\end{align}
where $l>0$ and $\mathbf{Ei}(x) = \int_{-\infty}^{x} \frac{e^t}{t} dt$.
Hence, the outage probability is given as 
\begin{align}
P(L < \gamma_{th}) =\int_{0}^{\gamma_{th}} f_L(l) dl.\label{out}
\end{align}
Consequently, based on the PDF of $L$ in~\eqref{pdf_pTpSIC}, the outage probability can be derived as \eqref{out_gamma_22k}. \qed
}

\textcolor{black}{
\section{Proof of Proposition 1}
Based on \eqref{sinr_k_STBC-CNOMA} and \eqref{Pout_k1}, the rate outage probability for the perfect timing synchronization, perfect SIC, and perfect CSI can be expressed as
\begin{align}
 \tilde{P}_{out} = {} &\mathop{\mathbb{P}}[\gamma_{k} < 2^{\Upsilon}-1]\nonumber\\
 =& \mathop{\mathbb{P}}\Bigg[[\frac{|h_{k}|^2 p_{k} }{|h_{k}|^2\sum\limits_{i=1}^{k-1}{ p_{i} + \sigma^2}} \nonumber\\
 &+ \frac{(|g_{k,k-\iota-1}|^2+|g_{k,k-\iota-2}|^2) {p}_{s} }{ \sigma^2}] < 2^{\Upsilon}-1\Bigg].\label{AsPout_knew}
 \end{align}
As described in Section IV, $|h_{k}|^2$, $|g_{k,k-\iota-2}|^2$, and $|g_{k,k-\iota-1}|^2$ follow the exponential distributions with parameters $\lambda_h$, $\lambda_{g1}$, and $\lambda_{g2}$, respectively. Therefore, we can rewrite \eqref{AsPout_knew} in terms of $\SNR$ as
 \begin{align}
  \tilde{P}_{out} = {} & \mathop{\mathbb{P}}\Big[[\tilde{X}+\tilde{Y} ] < g(\SNR) \Big], \label{AsPout_k2}
\end{align}
where $\tilde{X}=\frac{ |h_{k}|^2 \Phi_{k} }{\SNR |h_{k}|^2\sum\limits_{i=1}^{k-1}{ \Phi_{i} + 1}}$ and  $\tilde{Y}=\frac{|g_{k,k-\iota-1}|^2+|g_{k,k-\iota-2}|^2  }{ 2}$. 
 Because of the independent channel gains, $\mathop{\mathbb{P}}[\tilde{X}+\tilde{Y}< g(\SNR)]\leq \mathop{\mathbb{P}}[\tilde{X}< g(\SNR)] + \mathop{\mathbb{P}}[\tilde{Y}< g(\SNR)]$. At high $\SNR$, $\frac{\lambda_h(2^{\Upsilon}-1)}{\SNR[\Phi_k-\sum\limits_{i=1}^{k-1}\Phi_i( 2^{\Upsilon}-1)]}\rightarrow{0}$. Using the power series expansion~\cite{tailor1}, we first have 
  \begin{align}
  \mathop{\mathbb{P}}[\tilde{X}< g(\SNR)] =&  1-\exp\Bigg[-\frac{\lambda_h g(\SNR)}{\Phi_k-\sum\limits_{i=1}^{k-1}\Phi_i (2^{\Upsilon}-1)}\Bigg] \nonumber
  \end{align}
  \begin{align}
 &=1-\exp\Bigg[-\frac{\lambda_h(2^{\Upsilon}-1)}{\SNR\big(\Phi_k-\sum\limits_{i=1}^{k-1}\Phi_i( 2^{\Upsilon}-1)\big)}\Bigg]\nonumber\\
  &\thicksim \frac{\lambda_h}{\Phi_k-\sum\limits_{i=1}^{k-1}\Phi_i( 2^{\Upsilon}-1)}\Bigg(\frac{2^{\Upsilon}-1}{\SNR}\Bigg).\label{P_tid_x}
\end{align}
In addition, based on Fact-1 and Fact-2 in Appendix-I and Eq. (26) of~\cite{asymp2} to find the outage behavior when $\SNR\rightarrow{\infty}$, we obtain
  \begin{align}
  \mathop{\mathbb{P}}[\tilde{Y}< g(\SNR)] \thicksim 2\lambda_{g}^2\left(\frac{2^{\Upsilon}-1}{\SNR}\right)^2.\label{P_tid_y}
\end{align}
By substituting \eqref{P_tid_x} and \eqref{P_tid_y} into \eqref{AsPout_k2},
we have the asymptotic rate outage probability in \eqref{AsPout_k2f}. \qed
}

\textcolor{black}{
\section{Proof of Lemma 2}
For the imperfect SIC, perfect timing synchronization \textcolor{black}{and perfect CSI}, we can rewrite \eqref{SINR_n_pTim_ipSIC} as $L_\eta=\frac{A}{F+B}+Z$. Let $A$ and $B$  be the RVs as defined in~\eqref{pdf_a} and \eqref{pdf_b_i_m}. Also, suppose $F \sim Exp(\lambda_{\eta})$ and $\varpi=F+B$. Then, the PDF of $\varpi$ is given as
 \begin{align}
f_{\varpi}(\varpi) ={}&\lambda_\eta \prod_{i=1}^{I} \lambda_i\Bigg[ \sum\limits_{i=1}^{I}\Bigg(\frac{e^{-\lambda_i\varpi}}{ \prod_{j=1,i\neq j,\eta\neq j}^{I}(\lambda_j-\lambda_i)(\lambda_j-\lambda_\eta)} \nonumber\\
&+\frac{e^{-\lambda_\eta\varpi}}{ \prod_{j=1,i\neq j,\eta\neq j}^{I}(\lambda_j-\lambda_i)(\lambda_j-\lambda_\eta)}\Bigg)\Bigg], 
 \label{pdf_b_varpi}
\end{align}
where $\varpi>0$.
The PDF of  $Q_\eta=\frac{A}{\varpi}$ is given as
\begin{align}
f_{Q_\eta}(q) = \int_{0}^{\infty}\varpi f_A(\varpi q)f_{\varpi}(\varpi)d\varpi .
 \label{pdf_q_B_varpi}
\end{align}
\textcolor{black}{By substituting \eqref{pdf_a} and \eqref{pdf_b_varpi} into \eqref{pdf_q_B_varpi}, we obtain the PDF~\cite{prob1} of $Q_\eta=\frac{A}{F+B}$ is given as} 
\begin{align}
f_{Q_\eta}(q) &= \frac{ \psi^{\eta}_1\left[\psi^{\eta}_2 + q\lambda_h \Bigg( \sum\limits_{i=1}^{I}(i+2)q^{i}\lambda_h^{i} \psi^{I-i}_\eta\Bigg) \right]} {(\lambda_h q+\lambda_{\eta})^2 \prod_{j=1}^{I}  (\lambda_h q+\lambda_j)^2},
 \label{pdf_ipSIC}
\end{align}
where $q>0$. Thus, the mean and variance of $Q_\eta$ are given by
$\mathop{\E}[Q_\eta] = \frac{ \psi_{1}^\eta \psi_{3}^\eta} {\psi_{4}^\eta}$,
and 
$\mathop{\Var}[Q_\eta] = \frac{ \psi_{1}^\eta(\psi_{1}^\eta(\psi_{7}^\eta)^2+2 \psi_{8}^\eta\psi_{9}^\eta)} {\psi_{10}^\eta}$,
respectively. By convolving~\eqref{pdf_ipSIC} and ~\eqref{pdf_z}~\cite{prob1}, We get the PDF \textcolor{black}{\cite{prob1}} of $L_\eta=\frac{A}{F+B}+Z$ given as
\begin{align}
f_{L_\eta}(l)= {} &\sum\limits_{i=1}^{I}\Bigg[\frac{\mho e^{-\frac{\lambda_\eta +\lambda_i+\lambda_h l}{\lambda_g \lambda_h}}}{\lambda_g^3 \lambda_h^2 (\lambda_i-\lambda_\eta )} \Bigg(e^{\frac{\lambda_i}{\lambda_g
   \lambda_h}} \Bigg(-\lambda_\eta  \lambda_i \text{Ei}\left(\frac{\lambda_\eta }{\lambda_g \lambda_h}\right) \nonumber\\
{} & (-\lambda_g \lambda_h+\lambda_\eta +\lambda_h l) +\lambda_\eta  \lambda_i(-\lambda_g \lambda_h+\lambda_\eta +\lambda_h l)\nonumber\\
{} &\text{Ei}\left(\frac{l \lambda_h+\lambda_\eta }{\lambda_g \lambda_h}\right)+\lambda_g \lambda_h^2 l (\lambda_\eta
   -\lambda_i) \left(-e^{\frac{\lambda_\eta }{\lambda_g \lambda_h}}\right)\Bigg)\nonumber\\
{} &+\lambda_\eta  \lambda_i \left(-e^{\frac{\lambda_\eta
   }{\lambda_g \lambda_h}}\right)\text{Ei}\left(\frac{\lambda_i}{\lambda_g \lambda_h}\right) \bigg(\lambda_g \lambda_h\nonumber\\
{} &-\lambda_i+\lambda_h (-l)\bigg) +\lambda_\eta  \lambda_i e^{\frac{\lambda_\eta }{\lambda_g \lambda_h}} (\lambda_g
   \lambda_h-\lambda_i\nonumber\\
{} &+\lambda_h (-l)) \text{Ei}\left(\frac{l \lambda_h+\lambda_i}{\lambda_g \lambda_h}\right)\Bigg)\Bigg],
 \label{pdf_pTipSIC}
\end{align}
where $~l>0$. 
\textcolor{black}{By substituting \eqref{pdf_pTpSIC} into \eqref{out}, we get the outage probability~\cite{prob1} as shown in \eqref{out_gamma_k_eta}}. \qed
}

\begin{table}[t]
\caption{K-S Test Results for Lemma 3}
\label{Table3}
\centering
\begin{tabular} {|P{1.2cm}||P{.7cm}|P{.7cm}|P{.7cm}|P{.7cm}|P{.7cm}|P{.7cm}|}
\hline
\multirow{2}{*}{Distrns.} & 
\multicolumn{2}{c|}{Mean and Variance} & 
\multicolumn{2}{c|}{Estimated Para.} & 
\multicolumn{2}{c|}{MSE}\\
\cline{2-7}
 &$\mu$ & Var. & $\hat{\kappa_1}$ & $\hat{\kappa_2}$ &$e_{\kappa_1}$ &$e_{\kappa_2}$\\
\hline\hline
\rowcolor{Gray}
Gamma & 5.9834& 23.87 & 1.498& 3.9893&0.0061  &0.0192 \\
\hline
 Wei-bull& 6.0164&24.9705 &6.4089 & 1.2095&0.0177  &0.0028 \\
\hline
 Exponential&5.9834 &35.8037 & 5.9834& & 0.0189 & \\
\hline
 Rayleigh&7.2006 &14.1672 &5.7452 & &0.0092  & \\
\hline
 Rician&7.2010 &14.1687 &0.1753 & 5.7442&0.3146  &0.0102 \\
\hline
 Nakagami&6.4162 & 24.8484&0.4744 & 66.016&0.0017  & 0.3030\\
\hline
\end{tabular}
\end{table}

\textcolor{black}{
\section{Proof of Proposition 2}
Based on \eqref{SINR_n_pTim_ipSIC} and \eqref{Pout_k1}, the rate outage probability for the perfect timing synchronization, imperfect SIC, and perfect CSI can be rewritten as
\begin{align}
 \tilde{P}_{out}^{\eta} = {} &\mathop{\mathbb{P}}[\gamma_{k}^{\eta} < 2^{\Upsilon}-1]\nonumber\\
  =& \mathop{\mathbb{P}}\Bigg[\frac{ |h_{k}|^2 \Phi_{k} }{\SNR(\eta  |g_{\eta}|^2 \Phi_{\eta}+\sum\limits_{i=1}^{k-1}{|h_{k}|^2 \Phi_{i}) + 1}} \nonumber\\
 &+ \frac{|g_{k,k-\iota-1}|^2+|g_{k,k-\iota-2}|^2  }{ 2} < \frac{2^{\Upsilon}-1}{\SNR} \Bigg]. \label{AsPout_k_eta_1}
\end{align}
If $\tilde{Z}=\frac{ |h_{k}|^2 \Phi_{k} }{\SNR(\eta  |g_{\eta}|^2 \Phi_{\eta}+\sum\limits_{i=1}^{k-1}{|h_{k}|^2 \Phi_{i}) + 1}}$, using power series expansion, we have 
\begin{align}
  \mathop{\mathbb{P}}[\tilde{Z}< g(\SNR)] &={}1-\exp\Big[-\tilde{g}(\SNR)\Big]\thicksim \tilde{g}(\SNR).\label{P_tid_z}
\end{align}
\textcolor{black}{By substituting \eqref{P_tid_z} and \eqref{P_tid_y} into \eqref{AsPout_k_eta_1}, we can approximate the outage behaviour when $\SNR \rightarrow{\infty}$ as in \eqref{AsPout_k3f}}. \qed
}

\textcolor{black}{
\section{Proof of Lemma 3}
For the perfect SIC, imperfect timing synchronization, and perfect CSI, we can rewrite \eqref{SINR_k_even_ep} as $V=Q_1+R$. Further, the second part of \eqref{SINR_4_ep} can be expressed as $R = \frac{\upnu}{\Uplambda}$, where $\upnu=\frac{(C+\varepsilon_1 D)^2}{C+D}$ and $\Uplambda=\frac{|\aleph|^2}{C+D}$ with $|\aleph|^2=|(1-\varepsilon_1 )g_{4,1}g_{4,2}^*-\varepsilon_2g_{4,1}g_{4,2}^*|^2$. It can be shown from~\cite{GGR} that $\upnu$ is the Generalized Gamma distribution. Whereas, let $\Uplambda$ be a Gamma distribution. Then, the PDF of $\Uplambda$ is Gamma distribution~\cite{fourier_int}. Also, $R$ is a ratio of Generalized Gamma Distribution and Gamma Distribution~\cite{GGR}. Thus, the PDF of $R$ is given as in ~\eqref{pdf_z_ipTim}.
Similarly, suppose  $C \sim Exp(\lambda_{g})$, $D \sim Exp(\lambda_{g})$ , $0<\varepsilon_1\leq 1$, and $0<\varepsilon_2\leq 1$.  Then, the PDF of $R=\frac{(C+\varepsilon_1D)^2}{(1-\varepsilon_1)^2CD+\varepsilon_2^2CD+C+D}$ is 
\begin{align}
f_R(r) ={} &\frac{1}{4 (\varepsilon_1 -1)}\lambda_g\Bigg[\frac{1}{\sqrt{\lambda_g r}}\Bigg(\sqrt{\pi } \text{erf}\left(\sqrt{\lambda_g r}\right)\nonumber\\
{} &-\text{erf}\left(\frac{\sqrt{\lambda_g r}}{\varepsilon_1 }\right)\Bigg)+ \frac{2 e^{-\frac{\lambda_g r}{\varepsilon_1 ^2}}}{\varepsilon_1 }-2 e^{\lambda_g(-r)}\Bigg],
 \label{pdf_z_ipTim}
\end{align}
where $r>0$. Also, using this PDF, we can obtain the mean and variance as
\begin{align}
\mathop{\E}[R] = \frac{2( 1+\varepsilon_1 + \varepsilon_1^2)} {3\lambda_g},
 \label{mean_R}
\end{align}
and 
\begin{align}
\mathop{\Var}[R] = \frac{2 (17+7\varepsilon_1 -3\varepsilon_1^2 +7\varepsilon_1^3+17\varepsilon_1^4     )} {45\lambda_g^2},
 \label{var_R}
\end{align}
respectively. By applying the Kolmogorov–Smirnov test (K-S test)~\cite{k_S_test_book,K_S_test} on the distribution of $V=Q_1+R$, it is determined that $V \thicksim \Gamma (\alpha,\beta)$. Further, the PDF of $V=Q_1+R$ is given as
\begin{align}
f_V(v) = \frac{\beta ^{-\alpha } v^{\alpha -1} e^{-\frac{v}{\beta }}}{\Gamma (\alpha )},~v>0,
 \label{pdf_iT}
 \end{align}
where $\alpha= \frac{\mathop{\E}[V]^2}{\Var[V]}$ and $\beta=\frac{\Var[V]}{\mathop{\E}[V]}$ are the parameters of Gamma distribution.
Also, $\Gamma(x)$ is the gamma function given as $\Gamma(x)=(x-1)$!.
By applying the mathematical operation~\cite{prob1} and~\cite{prop_of_RV}, as given in~\eqref{out} on the PDF of $V$ in~\eqref{pdf_iT}, we obtain the outage probability as shown in \eqref{out_gamma_k_varepsilon}. \qed
}

\textcolor{black}{
\section{Proof of Proposition 3}
Based on \eqref{SINR_k_even_ep} and \eqref{Pout_k1}, the rate outage probability for the imperfect timing synchronization, perfect SIC, and perfect CSI can be expressed as
\begin{align}
 \tilde{P}_{out}^{\epsilon} = {} &\mathop{\mathbb{P}}[\gamma_{k}^\epsilon < 2^{\Upsilon}-1]\nonumber\\
 =& \mathop{\mathbb{P}}\Bigg[\frac{ |h_{k}|^2 \Phi_{k} }{\SNR\sum\limits_{i=1}^{k-1}{|h_{k}|^2 \Phi_{i} + 1}} \nonumber\\
 &+ \frac{(|\varphi_1|^2 + \varepsilon_1|\varphi_2|^2)^2}{{\SNR|\varphi_\varepsilon|^2}+{2(|\varphi_1|^2 + |\varphi_2|^2)}} < \frac{2^{\Upsilon}-1}{\SNR} \Bigg], \label{AsPout_k_epi_1}
\end{align}
where $\varphi_1=|g_{k,k-\iota-1}|^2$, $\varphi_2=|g_{k,k-\iota-2}|^2$, and $|\varphi_\varepsilon|^2 = |(1-\varepsilon_1 )g_{k,k-\iota-2}g_{k,k-\iota-1}^*-\varepsilon_2g_{k,k-\iota-2}g_{k,k-\iota-1}^*|^2$. Letting $\tilde{W}=\frac{ (|\varphi_1|^2 + \varepsilon_1|\varphi_2|^2)^2}{{\SNR|\varphi_\varepsilon|^2}+{2(|\varphi_1|^2 + |\varphi_2|^2)}}$, based on Fact-1 and Fact-2 in Appendix-I of~\cite{asymp2}, we can obtain
%
  \begin{align}
  \mathop{\mathbb{P}}[\tilde{W}< g(\SNR)] \thicksim g^{\varepsilon}(\SNR),\label{P_tid_w}
\end{align}
where $g^{\varepsilon}(\SNR)=2\varepsilon_1\lambda_{g}^2\left(\frac{2^{\Upsilon}-1}{\SNR}\right)^2$. By substituting \eqref{P_tid_w} and \eqref{P_tid_x} into \eqref{AsPout_k_epi_1}, we can approximate the outage behaviour when $\SNR \rightarrow{\infty}$ as in \eqref{AsPout_epif}. \qed
}

\textcolor{black}{
\section{Proof of Lemma 4}
For the imperfect SIC, imperfect timing synchronization, and perfect CSI, we first rewrite \eqref{SINR_k_even_ep_eta} as $V_\eta=Q_\eta+R$, where the PDFs of $Q_\eta$ and $R$ are given in~\eqref{pdf_ipSIC} and~\eqref{pdf_z_ipTim}, respectively. Applying the Kolmogorov–Smirnov test (K-S test)~\cite{k_S_test_book,K_S_test} on the distribution of $V_\eta=Q_\eta+R$, it is determined that $V_\eta \thicksim \Gamma (\theta,\phi)$, which corresponds to the following PDF
\begin{align}
f_{V_\eta}(v) =\frac{\phi ^{-\theta } \left(v\right)^{\theta -1} e^{-\frac{\left(v\right)}{\phi }}}{\Gamma (\theta )},
 \label{pdf_gamma_eta_epi}
\end{align}
where $v>0$. In addition, $\theta  = \frac{(\mathop{\E}[V_\eta])^2}{\Var[V_\eta]}$ and $\phi =\frac{\Var[V_\eta]}{\mathop{\E}[V_\eta]}$ are the parameters of Gamma distribution,  
%
where the mean and the variance of the $V_\eta$ are given by
\begin{align}
\mathop{\E}[V_\eta] & = \int_{0}^{\infty} v f_{V_\eta}(v) dv  =  \frac{\psi_1^\eta\psi_{3}^\eta}{\lambda_h\psi_{a}^\eta}+\frac{2 \left(\varepsilon_1 ^2+\varepsilon_1 +1\right)}{3 \lambda_g}
\end{align}
and 
\begin{align}
&\Var[V_\eta]  ={} \int_{0}^{\infty} v^ 2f_{V_\eta}(v) dV-\mathop{\E}[V_\eta]^2  \nonumber\\
&\hspace{-3pt}=\hspace{-3pt}
 \frac{\psi_1^\eta(\psi_1^\eta(\psi_3^\eta)^2+2\psi_{7}^\eta \psi_8^\eta )}{\psi_9^\eta} 
\hspace{-2pt}+
\hspace{-2pt}\frac{2 \left(17 \varepsilon_1 ^4+7 \varepsilon_1 ^3-3 \varepsilon_1 ^2+7 \varepsilon_1 +17\right)}{45
   \lambda_g^2},
\end{align}
respectively.
By substituting \eqref{pdf_gamma_eta_epi} into \eqref{out}, as in~\cite{prob1}, we can derive the outage probability as shown in \eqref{out_gamma_k_epi}.\qed 
}

%
\textcolor{black}{
 \section{Proof of Proposition 4}
With \eqref{SINR_k_even_ep_eta} and \eqref{Pout_k1}, the rate outage probability for  the imperfect timing synchronization, imperfect SIC, and perfect CSI can be found as
\eqref{SINR_k_even_ep_eta} is less than $\gamma_{th}$, which can be written as
\begin{align}
 \tilde{P}_{out}^{\eta,\epsilon} = {} &\mathop{\mathbb{P}}[\gamma_{k}^{\eta,\epsilon} < 2^{\Upsilon}-1]\nonumber\\
  =& \mathop{\mathbb{P}}\Bigg[[\frac{ |h_{k}|^2 \Phi_{k} }{\SNR(\eta  |g_{\eta}|^2 \Phi_{\eta}+\sum\limits_{i=1}^{k-1}{|h_{k}|^2 \Phi_{i}) + 1}} \nonumber\\
 &+\frac{ (|\varphi_1|^2 + \varepsilon_1|\varphi_2|^2)^2}{{\SNR|\varphi_\varepsilon|^2}+{2(|\varphi_1|^2 + |\varphi_2|^2)}}] < \frac{2^{\Upsilon}-1}{\SNR} \Bigg]. \label{AsPout_k5}
\end{align}
By substituting \eqref{P_tid_y} and \eqref{P_tid_w} into \eqref{AsPout_k5}, we can obtain \eqref{AsPout_epi_etaf}. \qed
}
%

\textcolor{black}{
 \section{Proof of Lemma 5}
For the perfect SIC, perfect timing synchronization, and imperfect CSI given in \eqref{SINR_k_even_chi}, let $A_\varrho=\varrho_{n,n-3}$, $A_g=g_{n,n-3}$, $B_\varrho= \varrho_{n,n-2}$, $B_g=g_{n,n-2}$, $C_\chi=|A_\chi|^2p_s$, and  $D_\chi=|B_\chi|^2p_s$, where $A_\chi=A_\varrho+\rho A_g$ and $B_\chi=B_\varrho+\rho B_g$. As described in Section III-C, $A_\varrho\sim CN(0,\sigma_{\varrho}^2)$, $A_g\sim CN(0,\sigma_{g}^2)$, $B_\varrho\sim CN(0,\sigma_{\varrho}^2)$, $B_g\sim CN(0,\sigma_{g}^2)$. Therefore, we can find out that $A_\chi \sim CN(0, \sigma_{\varrho}^2+\rho^2\sigma_{g}^2)$ and $B_\chi \sim CN(0, \sigma_{\varrho}^2+\rho^2\sigma_{g}^2)$. We model $A_\chi$ and $B_\chi$ as mutually independent complex Gaussian RVs with variance $\sigma_\chi^2$. Then, their magnitudes (i.e., $|A_\chi|$ and $|B_\chi|$) follow the Rayleigh distribution, and their squared magnitudes  (i.e., $|A_\chi|^2$ and $|B_\chi|^2$) follow the exponential distributions with the parameter $\lambda_\chi$, where $\lambda_\chi = 1/\sigma_\chi^2 = 1/(\sigma_{\varrho}^2+\rho^2\sigma_{g}^2)$~\cite{ipCSI2,ipCSI3}. The PDFs of $C_\chi$ and $D_\chi$ are given as
$f_{C_\chi}(c) =\frac{e^{-\frac{c}{\lambda _\chi}}}{\lambda _\chi},~c>0,$
and 
$f_{D_\chi}(d) =\frac{e^{-\frac{d}{\lambda _\chi}}}{\lambda _\chi},~d>0,$
respectively. If we define another variable $Z_\chi=C_\chi +D_\chi$, it follows a Gamma distribution, and its PDF is obtained by convolving $f_{C_\chi}(c)$ and $f_{D_\chi}(d)$ as
\begin{align}
f_{Z_\chi}(z) =\frac{z e^{-\frac{z}{\lambda _\chi}}}{\lambda _\chi^2},~z>0.
 \label{pdf_z_chi}
\end{align}
For the perfect SIC, perfect timings and imperfect CSI, \eqref{SINR_k_even_chi} can be written as $L_\chi=Q_1 + Z_\chi$ and the PDF of $L_\chi$ is obtained by convolving \eqref{pdf_q} and \eqref{pdf_z_chi}, which is given as
\begin{align}
f_{L_\chi}(l) = {} &\frac{1}{\lambda _\chi^3 \lambda_h^2}\sum\limits_{i=1}^{I}\Bigg[\mho e^{-\frac{\lambda_i+\lambda_h l}{\lambda _\chi \lambda_h}} \Bigg(\lambda_i (-\lambda _\chi
   \lambda_h+\lambda_i+\lambda_h l)\nonumber\\
{} & \left(\text{Ei}\left(\frac{l \lambda_h+\lambda_i}{\lambda _\chi
   \lambda_h}\right)-\text{Ei}\left(\frac{\lambda_i}{\lambda _\chi \lambda_h}\right)\right)+\nonumber\\
   &\lambda _\chi\lambda_h e^{\frac{\lambda_i}{\lambda _\chi \lambda_h}} \left(\lambda_i-\lambda_i
   e^{l/\lambda _\chi}+\lambda_h l\right)\Bigg)\Bigg],~l>0. \label{pdf_gamma_chi}
\end{align}
As a result, the corresponding outage probability is obtained as \eqref{out_gamma_k_chi}. \qed
}

\textcolor{black}{
\section{Proof of Proposition 5}
From \eqref{SINR_k_even_chi} and \eqref{Pout_k1}, the rate outage probability for the perfect timing synchronization, perfect SIC, and imperfect CSI can be expressed as
\begin{align}
 \tilde{P}_{out}^{\chi} &= {} \mathop{\mathbb{P}}[\gamma_{k}^\chi < 2^{\Upsilon}-1]\nonumber\\
 =& \mathop{\mathbb{P}}\Bigg[[\frac{ |h_{k}|^2 \Phi_{k} }{\SNR\sum\limits_{i=1}^{k-1}{|h_{k}|^2 \Phi_{i} + 1}} 
 + \frac{(|A_\chi|^2 + |B_\chi|^2)}{2}] < \frac{2^{\Upsilon}-1}{\SNR} \Bigg]. \label{AsPout_k6}
\end{align}
Following the same procedure as in \eqref{P_tid_y}, we obtain the PDF of $Z_{\chi}$ as
\begin{align}
  \mathop{\mathbb{P}}[\tilde{Z_\chi}< g(\SNR)] \thicksim 2\lambda_{\chi}^2\Bigg(\frac{2^{\Upsilon}-1}{\SNR}\Bigg)^2.\label{P_tid_Z_chi}
\end{align}
Replacing \eqref{P_tid_y} and \eqref{P_tid_Z_chi} into \eqref{AsPout_k6}, we have the asymptotic rate outage probability in \eqref{AsPout_k6f}. \qed} 


\end{appendices}
\ifCLASSOPTIONcaptionsoff
  \newpage
\fi

\bibliographystyle{IEEEtran}

\begin{thebibliography}{10}
\providecommand{\url}[1]{#1}
\csname url@rmstyle\endcsname
\providecommand{\newblock}{\relax}
\providecommand{\bibinfo}[2]{#2}
\providecommand\BIBentrySTDinterwordspacing{\spaceskip=0pt\relax}
\providecommand\BIBentryALTinterwordstretchfactor{4}
\providecommand\BIBentryALTinterwordspacing{\spaceskip=\fontdimen2\font plus
\BIBentryALTinterwordstretchfactor\fontdimen3\font minus
  \fontdimen4\font\relax}
\providecommand\BIBforeignlanguage[2]{{%
\expandafter\ifx\csname l@#1\endcsname\relax
\typeout{** WARNING: IEEEtran.bst: No hyphenation pattern has been}%
\typeout{** loaded for the language `#1'. Using the pattern for}%
\typeout{** the default language instead.}%
\else
\language=\csname l@#1\endcsname
\fi
#2}}

\bibitem{sur3}
Z.~{Ding}, X.~{Lei}, G.~K. {Karagiannidis}, R.~{Schober}, J.~{Yuan}, and V.~K.
  {Bhargava}, ``A survey on non-orthogonal multiple access for {5G} networks:
  Research challenges and future trends,'' \emph{IEEE J. on Sel. Commun.}, vol.~35, no.~10, pp. 2181--2195, Oct. 2017.

\bibitem{Survey2}
L.~{Dai}, B.~{Wang}, Z.~{Ding}, Z.~{Wang}, S.~{Chen}, and L.~{Hanzo}, ``A
  survey of non-orthogonal multiple access for {5G},'' \emph{IEEE
  Commun. Surveys Tutorials}, vol.~20, no.~3, pp. 2294--2323, 2018.

\bibitem{Survey3}
M.~{Vaezi}, G.~A. {Aruma Baduge}, Y.~{Liu}, A.~{Arafa}, F.~{Fang}, and
  Z.~{Ding}, ``Interplay between {NOMA} and other emerging technologies: A
  survey,'' \emph{IEEE Trans. Cognitive Commun. and
  Networking}, vol.~5, no.~4, pp. 900--919, Dec. 2019.

\bibitem{Survey4}
D.~{Wan}, M.~{Wen}, F.~{Ji}, H.~{Yu}, and F.~{Chen}, ``Non-orthogonal multiple
  access for cooperative communications: Challenges, opportunities, and
  trends,'' \emph{IEEE Wireless Commun.}, vol.~25, no.~2, pp. 109--117,
  Apr. 2018.

\bibitem{ZDCNOMA}
Z.~Ding, M.~Peng, and H.~V. Poor, ``Cooperative non-orthogonal multiple access
  in {5G} systems,'' \emph{IEEE Commun. Lett.}, vol.~19, no.~8, pp.
  1462--1465, Aug. 2015.

\bibitem{complx1}
Y.~Liu, Z.~Ding, M.~Elkashlan, and H.~V. Poor, ``Cooperative non-orthogonal
  multiple access with simultaneous wireless information and power transfer,''
  \emph{IEEE J. on Sel. Commun.}, vol.~34, no.~4, pp.
  938--953, 2016.

\bibitem{complx3}
C.~Zhong and Z.~Zhang, ``Non-orthogonal multiple access with cooperative
  full-duplex relaying,'' \emph{IEEE Commun. Lett.}, vol.~20, no.~12,
  pp. 2478--2481, 2016.

\bibitem{04}
Z.~Zhang, Z.~Ma, M.~Xiao, Z.~Ding, and P.~Fan, ``Full-duplex
  device-to-device-aided cooperative non-orthogonal multiple access,''
  \emph{IEEE Trans. Vehicular Tech.}, vol.~66, no.~5, pp.
  4467--4471, May 2017.

\bibitem{01}
Z.~Ding, H.~Dai, and H.~V. Poor, ``Relay selection for cooperative {NOMA},''
  \emph{IEEE Wireless Commun. Lett.}, vol.~5, no.~4, pp. 416--419, Aug.
  2016.

\bibitem{CRS-NOMA}
J.~{Kim} and I.~{Lee}, ``Capacity analysis of cooperative relaying systems
  using non-orthogonal multiple access,'' \emph{IEEE Commun. Lett.},
  vol.~19, no.~11, pp. 1949--1952, Nov. 2015.

\bibitem{liu2018decode}
H.~Liu, Z.~Ding, K.~J. Kim, K.~S. Kwak, and H.~V. Poor, ``Decode-and-forward
  relaying for cooperative NOMA systems with direct links,'' \emph{IEEE
  Trans. Wireless Commun.}, vol.~17, no.~12, pp. 8077--8093, Oct.
  2018.

\bibitem{03}
X.~Liang, Y.~Wu, D.~W.~K. Ng, Y.~Zuo, S.~Jin, and H.~Zhu, ``Outage performance
  for cooperative {NOMA} transmission with an {AF} relay,'' \emph{IEEE
  Commun. Lett.}, vol.~21, no.~11, pp. 2428--2431, Nov. 2017.

\bibitem{jamal2018efficient}
M.~N. Jamal, S.~A. Hassan, D.~N.~K. Jayakody, and J.~J. Rodrigues, ``Efficient
  nonorthogonal multiple access: Cooperative use of distributed space-time
  block coding,'' \emph{IEEE Vehicular Tech. Mag.}, vol.~13, no.~4,
  pp. 70--77, Dec. 2018.

\bibitem{NOMA_STBC}
M.~{Toka} and O.~{Kucur}, ``Non-orthogonal multiple access with Alamouti
  space–time block coding,'' \emph{IEEE Commun. Lett.}, vol.~22,
  no.~9, pp. 1954--1957, Sep. 2018.

\bibitem{kader2016cooperative}
M.~F. Kader and S.~Y. Shin, ``Cooperative relaying using space-time block coded
  non-orthogonal multiple access,'' \emph{IEEE Trans. Vehicular
  Tech.}, vol.~66, no.~7, pp. 5894--5903, Jul. 2016.

\bibitem{jamal2017new}
M.~N. Jamal, S.~A. Hassan, and D.~N.~K. Jayakody, ``A new approach to
  cooperative {NOMA} using distributed space time block coding,'' \emph{in Proc. 
  IEEE PIMRC}, pp. 1--5, Oct. 2017.

\bibitem{23}
M.~R. Avendi, S.~Poorkasmaei, and H.~Jafarkhani, ``Differential distributed
  space-time coding with imperfect synchronization,''  \emph{in Proc. IEEE GLOBECOM}, pp. 3186--3191, 2014.

\bibitem{24}
M.~Hussain and S.~A. Hassan, ``Analysis of bit error probability for imperfect
  timing synchronization in virtual {MISO} networks,'' \emph{in Proc. IFIP Wireless Days (WD)}, pp. 1--6, Nov. 2014.

\bibitem{Imperfect_SIC}
M.~R. {Usman}, A.~{Khan}, M.~A. {Usman}, Y.~S. {Jang}, and S.~Y. {Shin}, ``On
  the performance of perfect and imperfect sic in downlink non orthogonal
  multiple access ({NOMA}),'' \emph{in Proc. Int. Conf. on Smart
  Green Tech. in Electrical and Info. Sys. (ICSGTEIS)}, pp. 102--106, Oct. 2016.

\bibitem{ipCSI1}
H.~T. Cheng, H.~Mheidat, M.~Uysal, and T.~M. Lok, ``Distributed space-time
  block coding with imperfect channel estimation,'' \emph{in Proc. IEEE ICC}, pp. 583--587,  2005.

\bibitem{ipCSI2}
D.~Gu and C.~Leung, ``Performance analysis of transmit diversity scheme with
  imperfect channel estimation,'' \emph{Electronics Lett.}, vol.~39, no.~4,
  pp. 402--403, 2003.

\bibitem{ipCSI3}
J.~N. {Laneman}, ``Limiting analysis of outage probabilities for diversity
  schemes in fading channels,'' \emph{in Proc. IEEE GLOBECOM}, pp.
  1242--1246, 2003.

\bibitem{Alamouti}
S.~M. Alamouti, ``A simple transmit diversity technique for wireless
  communications,'' \emph{IEEE J. on Sel. Commun.},
  vol.~16, no.~8, pp. 1451--1458, Oct 1998.

\bibitem{stbc1}
G.~Ganesan and P.~Stoica, ``Space-time block codes: A maximum SNR approach,''
  \emph{IEEE Trans. Info. Theory}, vol.~47, no.~4, pp.
  1650--1656, 2001.

\bibitem{DSTBC_Uysal2007}
H.~{Mheidat} and M.~{Uysal}, ``Non-coherent and mismatched-coherent receivers
  for distributed STBCs with amplify-and-forward relaying,'' \emph{IEEE
  Trans. Wireless Commun.}, vol.~6, no.~11, pp. 4060--4070,
  2007.

\bibitem{complx2}
L.~Lv, J.~Chen, and Q.~Ni, ``Cooperative non-orthogonal multiple access in
  cognitive radio,'' \emph{IEEE Commun. Lett.}, vol.~20, no.~10, pp.
  2059--2062, 2016.

\bibitem{complx4}
Z.~Zhang, Z.~Ma, M.~Xiao, Z.~Ding, and P.~Fan, ``Full-duplex
  device-to-device-aided cooperative nonorthogonal multiple access,''
  \emph{IEEE Trans. Vehicular Tech.}, vol.~66, no.~5, pp.
  4467--4471, 2016.

\bibitem{complx5}
L.~Lv, J.~Chen, Q.~Ni, and Z.~Ding, ``Design of cooperative non-orthogonal
  multicast cognitive multiple access for {5G} systems: User scheduling and
  performance analysis,'' \emph{IEEE Trans. Commun.}, vol.~65,
  no.~6, pp. 2641--2656, 2017.

\bibitem{CRS-NOMA-ND}
M.~{Xu}, F.~{Ji}, M.~{Wen}, and W.~{Duan}, ``Novel receiver design for the
  cooperative relaying system with non-orthogonal multiple access,'' \emph{IEEE
  Commun. Lett.}, vol.~20, no.~8, pp. 1679--1682, Aug. 2016.

\bibitem{prob1}
R.~E. Walpole, R.~H. Myers, S.~L. Myers, and K.~Ye, \emph{Probability and
  statistics for engineers and scientists}.\hskip 1em plus 0.5em minus
  0.4em\relax Macmillan New York, 1993, vol.~5.

\bibitem{tailor1}
I.~S. Gradshteyn and I.~M. Ryzhik, \emph{Table of integrals, series, and
  products}.\hskip 1em plus 0.5em minus 0.4em\relax Academic press, 2014.

\bibitem{asymp2}
J.~N. Laneman, D.~N. Tse, and G.~W. Wornell, ``Cooperative diversity in
  wireless networks: Efficient protocols and outage behavior,'' \emph{IEEE
  Trans. Info. Theory}, vol.~50, no.~12, pp. 3062--3080, 2004.

\bibitem{GGR}
\BIBentryALTinterwordspacing
C.~A. Coelho and J.~T. Mexia, ``On the distribution of the product and ratio of
  independent generalized gamma-ratio random variables,'' \emph{Sankhyā: The
  Indian Journal of Statistics (2003-2007)}, vol.~69, no.~2, pp. 221--255,
  2007. [Online]. Available: \url{http://www.jstor.org/stable/25664553}
\BIBentrySTDinterwordspacing

\bibitem{fourier_int}
N.~Wiener, \emph{The Fourier integral and certain of its applications}.\hskip
  1em plus 0.5em minus 0.4em\relax CUP Archive, 1988.

\bibitem{k_S_test_book}
I.~M. Chakravarti, R.~G. Laha, and J.~Roy, ``Handbook of methods of applied
  statistics,'' \emph{Wiley Series in Probability and Mathematical Statistics
  (USA) eng}, 1967.

\bibitem{K_S_test}
A.~G. {Glen}, L.~M. {Leemis}, and D.~R. {Barr}, ``Order statistics in
  goodness-of-fit testing,'' \emph{IEEE Trans. Reliability}, vol.~50,
  no.~2, pp. 209--213, Jun. 2001.

\bibitem{prop_of_RV}
\BIBentryALTinterwordspacing
C.~Bettstetter, H.~Hartenstein, and X.~P{\'e}rez-Costa, ``Stochastic properties
  of the random waypoint mobility model,'' \emph{Wireless Networks}, vol.~10,
  no.~5, pp. 555--567, Sep 2004. [Online]. Available:
  \url{https://doi.org/10.1023/B:WINE.0000036458.88990.e5}
\BIBentrySTDinterwordspacing

\end{thebibliography}
\end{document}